\documentclass[11pt]{article}

\usepackage{cite}
\usepackage{hyperref}
\usepackage{soul}

\usepackage[dvipsnames]{xcolor}
\usepackage[normalem]{ulem}

\usepackage{lineno}

\usepackage{amsmath}
\usepackage{bbold}
\usepackage{amssymb,amsfonts,amsthm}
\usepackage{epsfig,graphicx}
\usepackage{url} 
\usepackage{bm} 
\usepackage{slashed}
\usepackage{cite}

\usepackage{color}
\usepackage{pstricks}

\usepackage{fancyhdr}
\usepackage{textcomp}

\usepackage{float}
\usepackage{caption}
\usepackage{subcaption}
\usepackage{enumitem}

\usepackage{feynmp}
\def\MET{E_{\rm T} \hspace{-1.2em}/\;\;}

\def\MET{E_{\rm T} \hspace{-1.2em}/\;\;}

\newcommand{\Real}{\text{Re}}
\newcommand{\Imag}{\text{Im}}

\def \GeV{{\mathrm{GeV}}}
\def \TeV{{\mathrm{TeV}}}

\DeclareMathAlphabet{\pazocal}{OMS}{zplm}{m}{n}
\newcommand{\Ha}{\pazocal{H}}
\newcommand{\La}{\pazocal{L}}
\numberwithin{equation}{section}

\def\bsmumu{b \rightarrow s\mu^{+}\mu^{-}}
\DeclareMathOperator\arctanh{arctanh}

\def\mchi{m_{\chi}}
\def\mphil{m_{\phi_l}}
\def\mphiq{m_{\phi_q}}
\def\lambdas{\lambda_s}
\def\lambdab{\lambda_b}
\def\lambdamu{\lambda_\mu}
\def\sigmav{\langle\sigma v\rangle}

\begin{document}

\begin{flushright}
IPPP/19/8\\
IFT-UAM/CSIC-19-13
\end{flushright}

\begin{center}

  {\bf {\LARGE B anomalies and dark matter: a complex connection}}
  
\renewcommand*{\thefootnote}{\fnsymbol{footnote}}
\setcounter{footnote}{0}

  \vspace{0.5cm}
  {\large
    D.~G.~Cerde\~no $^{a}$,
    A.~Cheek $^{a}$, 
    P.~Mart\'\i n-Ramiro $^{b}$, and
    J.~M.~Moreno $^{b}$
  }
  \\[0.2cm] 

  {\footnotesize{
$^a$ Institute for Particle Physics Phenomenology, Department of Physics\\
Durham University, Durham DH1 3LE, United Kingdom \\[1ex]
$^b$ Instituto de F\'\i sica Te\'orica, IFT-UAM/CSIC \\ Universidad Aut\'onoma de Madrid, Madrid 28049, Spain
        }
    }

\vspace*{0.7cm}

\begin{abstract}
We study an extension of the Standard Model that addresses the hints of  lepton flavour universality violation observed in $B\to K^{(*)} l^+l^-$ decays at LHCb, while providing a viable candidate for dark matter.  The model incorporates two new scalar fields and a Majorana fermion that induce one-loop contributions to $B$ meson decays. We show that agreement with observational data requires the new couplings to be complex and that the Majorana fermion can reproduce the observed dark matter relic density. This combination of cosmological and flavour constraints sets an upper limit on the dark matter and mediator masses. We have studied LHC dijet and dilepton searches, finding that they rule out large regions of parameter space by setting lower bounds on the dark matter and mediator masses. In particular, dilepton bounds are much more constraining in a future high-luminosity phase. Finally, we have computed the scattering cross section of dark matter off nuclei and compared it to the sensitivity of current and future direct detection experiments, showing that parts of the parameter space could be accessible in the future to multi-ton experiments. Future collider and direct DM searches complement each other to probe large areas of the parameter space of this model.
\end{abstract}
\end{center}

\newpage
\section{Introduction}
\label{sec:Intro}

LHCb has reported anomalies in the measured decay rates of the $B$ meson, which have been interpreted as hints of lepton flavour universality violation \cite{Aaij:2014pli,Aaij:2015esa}. The SM predicts equal rates for the processes $B\to K^{(*)} \mu^+\mu^-$ and $B\to K^{(*)} e^+e^-$, and it is customary to study the ratios of these branching ratios, defined as $R(K)$ and $R(K^*)$, since the dependencies on hadronic matrix elements (and associated uncertainties) cancel out \cite{Hiller:2003js}. The measurements of these hadronically clean observables deviate consistently (although perhaps with not enough statistical significance) from the SM prediction $R(K^{(*)})=1$ \cite{Bordone:2016gaq}. These hints are complemented by measurements of other observables that are more sensitive to hadronic physics. In particular, the differential branching fractions \cite{Wei:2009zv,Aaij:2014pli,Aaij:2015esa} and angular observables \cite{Khachatryan:2015isa,Abdesselam:2016llu,Lees:2015ymt,Sirunyan:2017dhj,Aaboud:2018krd,Wehle:2016yoi,Aaltonen:2011ja,Aaij:2015oid} associated to the processes $B\to \phi \mu^+\mu^-$ and $B\to K^{(*)} \mu^+\mu^-$ also deviate from the SM predictions. Interestingly, all the apparent anomalies involve the transition $\bsmumu$.

In order to account for these experimental results, one can modify the SM effective Hamiltonian, which involves penguin and box diagrams, by including one-loop contributions from new exotic particles. A full classification of the various particle combinations, considering different gauge representations, was presented in  Refs.~\cite{Gripaios:2015gra,Arnan:2016cpy}. Among the different models, some featured neutral scalar or fermions that, if stable, could play the role of dark matter (DM)\footnote{An alternative to this one-loop solution is to consider $ Z^\prime$~\cite{Buras:2013qja,Gauld:2013qja} or leptoquark~\cite{Bauer:2015knc,Angelescu:2018tyl} tree-level contributions, see e.g., Ref.~\cite{Capdevila:2017bsm} and references therein. The DM problem has been addressed in the framework of these constructions \cite{Vicente:2018xbv}, see e.g.,  Refs.~\cite{Sierra:2015fma,Belanger:2015nma,Altmannshofer:2016jzy,Celis:2016ayl,Cline:2017lvv,Ko:2017yrd,Ellis:2017nrp,Baek:2017sew,Fuyuto:2017sys,Cox:2017rgn,Falkowski:2018dsl,Darme:2018hqg,Singirala:2018mio,Baek:2018aru,Kamada:2018kmi} for the $Z^\prime$, and Refs.~\cite{Hati:2018fzc,Choi:2018stw,Cline:2017aed,Varzielas:2015sno} for the leptoquark models.}. The first possibility was investigated in Ref.~\cite{Kawamura:2017ecz}, where it was found that the large new couplings required to reproduce the correct DM relic abundance induce sizeable 1-loop contributions to DM-nucleon scattering, leading to very strong limits from direct detection experiments. In addition, as reported by \cite{Bhattacharya:2015xha}, the Higgs portal coupling typically dominates over other new physics effects. The second possibility was addressed in Ref.~\cite{Cline:2017qqu}, where the fermionic dark matter field was accompanied by one additional scalar and one additional coloured fermion.

In this work, we consider a modification of the model of Ref.~\cite{Cline:2017qqu}. Namely, we will also assume a fermionic dark matter particle, but with two extra scalar fields, one of which has a colour charge. On top of this, we include the latest SM theoretical prediction for the mass difference in $B_{s}-$mixing \cite{DiLuzio:2017fdq}, which differs from the experimental observation by $1.8\,\sigma$. In order to reduce this tension and provide an explanation for the $B$ anomalies, complex couplings are needed, leading to new CP-violation sources, a scenario that has not been studied in the context of one-loop models so far. We explore the parameter space of this model, taking into account all the flavour observables, DM constraints, and LHC collider signatures.

This paper is organised as follows. In Section~\ref{sec:model}, we introduce the details of the particle physics model, address the constraints from the observed DM relic abundance and $B_{s}-$mixing and discuss the implications on the model's parameter space. In Section~\ref{sec:lhc}, we investigate the possibility of observing this scenario at the LHC, for which we take into account dijet and dimuon searches. We also include a projection of the potential reach of the High Luminosity phase of the LHC.  Finally, in Section~\ref{sec:direct}, we compute the DM-nucleus scattering cross section and study current constraints and the future reach of direct DM detection experiments. The conclusions are presented in Section~\ref{sec:conclusions}.

\section{The model}
\label{sec:model}

In this article, we consider a model in which the DM particle is a Majorana fermion, $\chi$, with two extra scalar fields, $\phi_q$ and $\phi_l$, which couple to left-handed quarks and leptons, respectively\footnote{As we will comment in Section \ref{sec:direct}, the alternative construction with Dirac DM is ruled out mainly by experimental results from direct DM detection.}. The interactions between the new particles and the SM are described by the Lagrangian,
\begin{equation}
  \La_{\text{int}}^{\text{NP}} = \lambda_{Q_i}\bar{Q}_{i}\phi_{q}P_{R}\chi + \lambda_{L_i}\bar{L}_{i}\phi_{l}P_{R}\chi + \rm{h.c.} \, ,
  \label{eq:lagrangian}
\end{equation}
where $Q_i$ and $L_i$ denote the SM left-handed quark and lepton doublets of each generation, and $\lambda_{Q_i}$ and $\lambda_{L_i}$ are the corresponding new couplings. The quantum numbers for the new fields are summarised in Table~\ref{tab:quantum_numbers}. We impose a ${\cal Z}_2$ parity under which the SM fields are invariant, and which guarantees the stability of the DM candidate, as long as $m_{\phi_{q,l}} > \mchi$. Upon rotation from the electroweak to the quark mass eigenbasis, the couplings $\lambda_{Q_i}$ are rotated in flavour space. Assuming that the electroweak and mass eigenbasis are aligned for the leptons and down-type quarks, the couplings to the up-type quarks are generated by the CKM rotation as follows:
\begin{equation}
  \lambda_{Q_i}\bar{Q}_{i} \rightarrow \lambda_{Q_j}(\bar{u}_{L,i}V_{ij}, \bar{d}_{L,j}) \, .
\label{eq:rotation}
\end{equation}
From now on, we will denote the couplings in the mass eigenbasis with the corresponding quark or lepton label. These couplings are, in general, complex.

\begin{table}[t!]
\begin{center}
\begin{tabular}{|c|c|c|c||c|}
\hline  
& $SU(3)$ & $SU(2)_{L}$ & $U(1)_{Y}$ & ${\cal Z}_2$ \\
\hline
\hline $\phi_{q}$ & $3$ & $2$ & $1/6$ & $-1$ \\
\hline $\phi_{l}$ & $1$ & $2$ & $-1/2$ & $-1$ \\
\hline $\chi$ & $1$ & $1$ & $0$ & $-1$ \\
\hline
\end{tabular}
\caption{Quantum numbers of the new fields. We also indicate the charges under ${\cal Z}_2$.}
\label{tab:quantum_numbers}
\end{center}
\end{table}

\begin{figure}[!t]
    \begin{center}
	\includegraphics[scale=0.8]{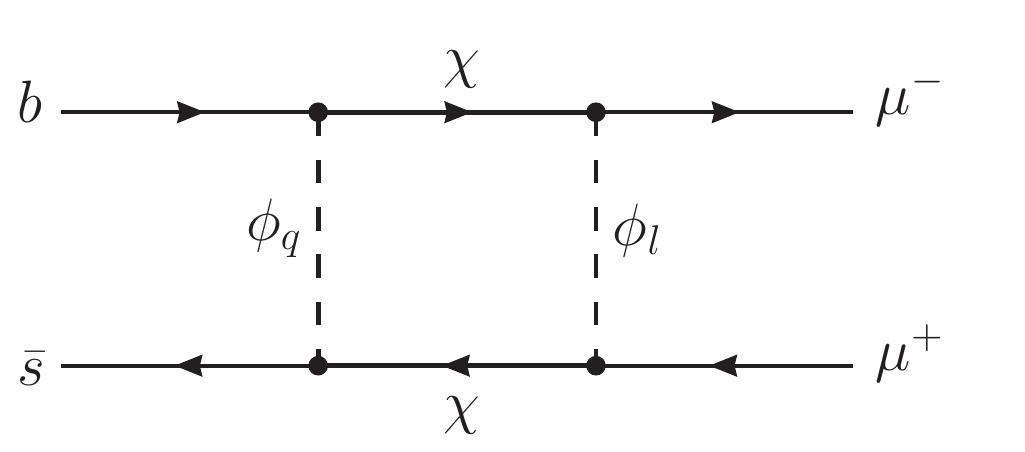}
    \end{center}
	\caption{One-loop diagram contribution from the new particles to the $\bsmumu$ transitions.}
    \label{fig:box}
\end{figure}

This model induces new physics contribution to flavour observables at the one loop level. In particular, a new box diagram appears for the  $\bsmumu$ transition, as shown in Figure~\ref{fig:box}. These effects  can be described using an effective field theory approach, thus parameterising the new contributions as corrections to the corresponding Wilson coefficients of the effective Hamiltonian,
\begin{equation}
  \Ha_{\text{eff}}^{\mu^{+}\mu^{-}} = -\frac{4G_{F}}{\sqrt{2}}V_{tb}V_{ts}^{*}(C_{9}\pazocal{O}_{9} + C_{9}^{\prime}\pazocal{O}_{9}^{\prime} + C_{10}\pazocal{O}_{10} + C_{10}^{\prime}\pazocal{O}_{10}^{\prime}) + \text{h.c.} \, ,
  \label{eq:hamiltonian}
\end{equation}
where the effective operators $\pazocal{O}_{9}$, $\pazocal{O}_{9}^{\prime}$, $\pazocal{O}_{10}$, $\pazocal{O}_{10}^{\prime}$ are defined as:
\begin{eqnarray}
  \pazocal{O}_{9} &=& \frac{\alpha_{\text{em}}}{4\pi}(\bar{s}\gamma^{\mu}P_{L}b)(\bar{\mu}\gamma_{\mu}\mu) \, , \\
  \pazocal{O}_{9}^{\prime} &=& \frac{\alpha_{\text{em}}}{4\pi}(\bar{s}\gamma^{\mu}P_{R}b)(\bar{\mu}\gamma_{\mu}\mu) \, ,\\
  \pazocal{O}_{10} &=& \frac{\alpha_{\text{em}}}{4\pi}(\bar{s}\gamma^{\mu}P_{L}b)(\bar{\mu}\gamma_{\mu}\gamma_{5}\mu) \, , \\
  \pazocal{O}_{10}^{\prime} &=& \frac{\alpha_{\text{em}}}{4\pi}(\bar{s}\gamma^{\mu}P_{R}b)(\bar{\mu}\gamma_{\mu}\gamma_{5}\mu) \, .
\end{eqnarray}
The Wilson coefficients $C_{9}$, $C_{9}^{\prime}$, $C_{10}$, $C_{10}^{\prime}$ contain both the SM and new physics (NP) contributions,
\begin{eqnarray}
  C_{9} &=&C_{9}^{\text{SM}} + C_{9}^{\text{NP}} \, ,\nonumber \\
  C_{10} &=& C_{10}^{\text{SM}} + C_{10}^{\text{NP}} \, ,
  \label{eq:c9c10}
\end{eqnarray}
with the primed coefficients defined in an equivalent way.

Global fits \cite{Descotes-Genon:2015uva,Hurth:2016fbr,Capdevila:2017bsm,Altmannshofer:2017yso,DAmico:2017mtc,Hiller:2017bzc,Geng:2017svp,Ciuchini:2017mik,Alok:2017sui,Hurth:2017hxg} have been used to determine the new physics contribution to the Wilson coefficients in order to reproduce the observed experimental results. These fits favour $C_{9}^{\text{NP}} = -C_{10}^{\text{NP}}$, and suggest that no new physics is required for operators involving electrons or tau leptons. Motivated by these results, we assume negligible couplings to the first quark generation (i.e., $\lambda_{Q_1} = 0$) and to the first and third lepton generations (i.e., $\lambda_{e}=\lambda_{\tau}=0$). This provides an explanation for the $R_{K^{(*)}}$ anomalies, while relaxing the bounds from other searches.

Therefore, in total, we are left with six free parameters in this model, namely the masses of the three new particles ($\mchi$, $\mphil$, $\mphiq$), and the couplings to $b-$type quarks, $s-$type quarks, and leptons ($\lambdab$, $\lambdas$, $\lambdamu$).

It should be noted that the couplings $\lambda_{1}|\phi_{l}|^{2}|H|^{2}$ and $\lambda_{2}|\phi_{q}|^{2}|H|^{2}$ are allowed by gauge symmetry in the Lagrangian of Eq.~(\ref{eq:lagrangian}). However, they only lead to an overall shift to the masses of $\phi_{l}$ and $\phi_{q}$ after electroweak symmetry breaking since the couplings to the Higgs play no phenomenological role in the relevant range of $\phi_{l,q}$ masses. Likewise, the terms $\lambda_{3}|\phi_{l}H|^{2}$ and $\lambda_{4}|\phi_{q}H|^{2}$ are also allowed by gauge symmetry. They typically induce a small split in the masses of the neutral and charged components of the doublets $\phi_{l}$ and $\phi_{q}$ in the range of $\phi_{l,q}$ masses that survive the collider constraints. Finally, a term of the form $(\phi_{l}H)^{2}$ can lead to large contributions to neutrino masses at one loop, which forces the corresponding coupling to be extremely small \cite{Cline:2017qqu}. We will neglect these couplings in the following.

As mentioned in the Introduction, similar models have been discussed in the literature, featuring either scalar DM \cite{Kawamura:2017ecz,Chiang:2017zkh,Barman:2018jhz,Grinstein:2018fgb} or fermionic DM \cite{Cline:2017qqu}. Our model differs from that of Ref.~\cite{Cline:2017qqu} in that we have two extra scalar fields which couple to the lepton or quark sectors.

\subsection{Dark matter relic abundance}
\label{sec:relic}

\begin{figure}[!t]
    \begin{center}
    \includegraphics[scale=0.82]{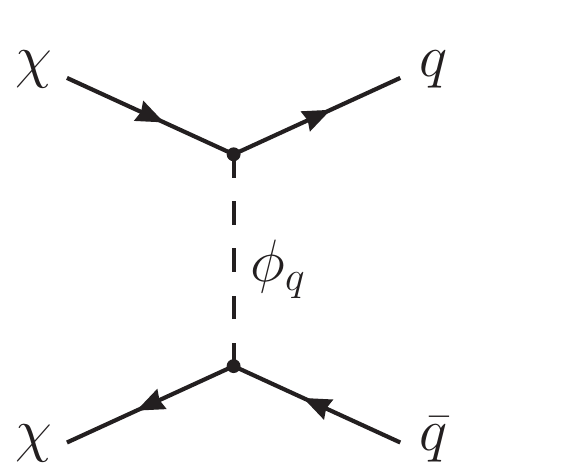}
	\includegraphics[scale=0.82]{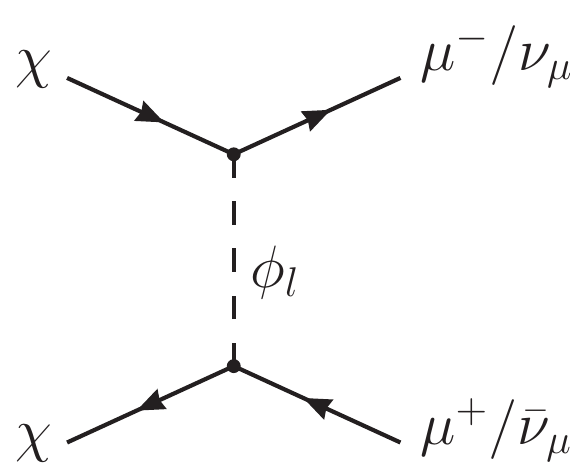}
    \end{center}
	\caption{Tree-level contributions to the DM pair annihilation.}
    \label{fig:annihilation}
\end{figure}

In order for $\chi$ to be a viable DM candidate, it must reproduce the observed relic abundance, which can be inferred from Planck satellite data to be $\Omega h^2=0.1199\pm 0.0022$ \cite{Ade:2015xua}. The pair-annihilation proceeds through the two $t-$channel diagrams with $\phi_q$ and $\phi_l$, shown in Figure~\ref{fig:annihilation}.

The stringent flavour constraints force the couplings to quarks to be much smaller than the couplings to leptons (muons and neutrinos), and the combination of flavour and collider bounds impose $\mphiq>\mphil$, with coloured scalars generally above 1~TeV. Therefore DM annihilation into a $\mu^-\mu^+$ or $\nu_\mu \bar\nu_\mu$ pair is the dominant channel. The thermally-averaged annihilation cross section, $\sigmav$, can be expressed as a plane wave expansion in terms of the dimensionless parameter $x=\mchi/T$. For the case of a Majorana fermion, the zero-velocity term is helicity suppressed, and the leading contribution comes from the linear term in $1/x$ \cite{Chang:2014tea},
\begin{equation}
  \sigmav = 2 \frac{|\lambda_{\mu}|^{4}m_{\chi}^{2}\left(m_{\phi_{l}}^{4} + m_{\chi}^{4}\right)}{16\pi \left(m_{\phi_{l}}^{2} + m_{\chi}^{2}\right)^{4}} \, \frac{1}{x}\ ,
  \label{eq:sigmav}
\end{equation}
where we have neglected the muon and the neutrino masses. In order to reproduce the correct relic abundance, we can now impose $\sigmav = 2.2\times 10^{-26}$~cm$^3$~s$^{-1}$ (where $x\sim 20$ at freeze-out).

We will use this relation to fix $\mphil$ as a function of the other parameters, thus effectively reducing by one the number of free parameters. Furthermore, due to the suppression of the velocity-independent term for $\sigmav$, indirect detection bounds are not expected to constrain our model.

\subsection{$B_{s}-$mixing and other flavour constraints}
\label{sec:bsmixing}

This model introduces new couplings to the $s$ and $b$ quarks (and to the rest of the quarks by rotation of the CKM matrix). We must therefore incorporate constraints from $B$ meson physics.

The most relevant bounds are those that involve $\bsmumu$ transitions. The new physics contribution to the Wilson coefficient comes from box and photon-penguin diagrams \cite{Gripaios:2015gra,Arnan:2016cpy}, $C_{9}^{\rm NP}=C_{9}^{\text{box}}+C_{9}^{\gamma}$, with\footnote{We have neglected the $Z$-penguin contribution to $C_{9}^{\text{NP}}$, since it is suppressed by $(m_{b}/m_{Z})^{2}$ and is subdominant compared to the photon exchange.}
\begin{eqnarray}
  C_{9}^{\text{box}} &=& \frac{\sqrt{2}}{128\pi\alpha_{\text{em}}G_{F}m^{2}_{\psi}} \frac{\lambda_{s}\lambda_{b}^{*}}{V_{tb}V_{ts}^{*}} |\lambda_{\mu}|^{2} \left(F(x_{q},x_{l}) + 2G(x_{q},x_{l}) \right) \, ,\nonumber\\
  C_{9}^{\gamma} &=& \frac{\sqrt{2}}{8G_{F}m^{2}_{\psi}} \frac{\lambda_{s}\lambda_{b}^{*}}{V_{tb}V_{ts}^{*}} F_{9}(x_{q}) \, ,
\end{eqnarray}
where we have defined the dimensionless variables $x_{q} = m_{\phi_{q}}^{2} / m_{\chi}^{2}$ and $x_{l} = m_{\phi_{l}}^{2} / m_{\chi}^{2}$, and the loop functions are:
\begin{eqnarray}
  F(x,y) &=& \frac{1}{(1-x)(1-y)} + \frac{x^{2}\log x}{(1-x)^{2}(x-y)} + \frac{y^{2}\log y}{(1-y)^{2}(y-x)} \, ,\nonumber\\
  G(x,y) &=& \frac{1}{(1-x)(1-y)} + \frac{x\log x}{(1-x)^{2}(x-y)} + \frac{y\log y}{(1-y)^{2}(y-x)} \, ,\nonumber\\
  F_{9}(x) &=& \frac{-2x^{3} + 9x^{2} - 18x + 6\log x + 11}{36(x-1)^{4}} \, .
\label{eq:fg}
\end{eqnarray}
The term  $G(x_{q},x_{l})$ vanishes if $\chi$ is a Dirac particle.

In order to constrain the Wilson coefficients we use the first global fit that takes into account the possibility that $C_{9}$ and $C_{10}$ are complex \cite{Alok:2017jgr}. This is a natural scenario that arises when new CP-violation sources are introduced, and has not been studied in detail in the literature so far.

Likewise, the new physics contribution to $B_{s}-$mixing can be parameterised in terms of an effective Hamiltonian,
\begin{equation}\label{eq:3.2.1}
  \Ha_{\text{eff}}^{b\bar{s}} = C_{B\bar{B}}^{\text{NP}} \, (\bar{s}_{\alpha}\gamma^{\mu}P_{L}b_{\alpha})(\bar{s}_{\beta}\gamma_{\mu}P_{L}b_{\beta}) \, ,
\end{equation}
where $\alpha$ and $\beta$ are colour indices. The new physics contribution to the Wilson coefficient is given by
\begin{equation}\label{eq:wilsonCBB}
  C_{B\bar{B}}^{\text{NP}} = \frac{1}{128\pi^{2}m^{2}_{\psi}} (\lambda_{s}\lambda_{b}^{*})^{2} \left(F(x_{q},x_{q}) + 2G(x_{q},x_{q}) \right) \, ,
\end{equation}
where the loop functions $F$ and $G$ were already defined in Eq.~(\ref{eq:fg}).

In order to quantify the allowed magnitude of the Wilson coefficient $C_{B\bar{B}}^{\text{NP}}$, we follow the steps of \cite{DiLuzio:2017fdq} and introduce a complex parameter $\Delta$ in the following way:
\begin{equation}\label{eq:3.2.3}
  \frac{M_{12}^{\text{SM}} + M_{12}^{\text{NP}}}{M_{12}^{\text{SM}}} \equiv |\Delta|e^{i\phi_{\Delta}} \, ,
\end{equation}
where $M_{12}^{\text{SM}}$ and $M_{12}^{\text{NP}}$ describe the SM and new physics contributions to $B_{s}-$mixing, and their values are given by the corresponding box diagrams. The complex phase, $\phi_{\Delta}$, quantifies the CP-violating effects introduced by the imaginary parts of the new couplings. We find:
\begin{eqnarray}
  |\Delta| &=& \frac{\Delta M_{s}^{\text{exp}}}{\Delta M_{s}^{\text{SM}}} = \left|1+\frac{C_{B\bar{B}}^{\text{NP}}}{C_{B\bar{B}}^{\text{SM}}} \right| \, ,\nonumber\\
  \phi_{\Delta} &=& \text{Arg} \left(1+\frac{C_{B\bar{B}}^{\text{NP}}}{C_{B\bar{B}}^{\text{SM}}} \right) \, ,
  \label{eq:deltadef}
\end{eqnarray}
where $\Delta M_{s}$ is the mass difference of the mass eigenstates of the $B_s$ meson.

The parameter $|\Delta|$ can be constrained using the most precise experimental measurement of $\Delta M_{s}$\cite{Amhis:2016xyh} and the last update on its theoretical prediction \cite{DiLuzio:2017fdq}, which show a $1.8 \sigma$ difference,
\begin{eqnarray}
\Delta M_{s}^{\text{exp}} &=& (17.757 \pm 0.021) \, \text{ps}^{-1} \, ,\nonumber\\
\Delta M_{s}^{\text{SM}} &=& (20.01 \pm 1.25) \, \text{ps}^{-1} \, .
\label{eq:deltams}
\end{eqnarray} 
The dominant uncertainties in the calculation of $\Delta M_{s}^{\text{SM}}$ come from lattice predictions for the non-perturbative bag parameter, $\mathcal{B}$, and decay constant, $f_{B_{s}}$, and to a lesser extent from the uncertainty in the values of CKM elements. Both of these errors have been considerably reduced since the last theory update for the mass difference \cite{Artuso:2015swg}. The last average given by the lattice community \cite{Aoki:2016frl} gives significantly more precise values for $\mathcal{B}$ and $f_{B_{s}}$.

From these values, one can infer $|\Delta| = 0.887 \pm 0.055$, and using the data provided in Ref.~\cite{DiLuzio:2017fdq} we obtain $C_{B\bar{B}}^{\text{SM}} = 4.897 \times 10^{-5} \; \TeV^{-2} \,$. Using Eq. (\ref{eq:deltadef}) we find that the Wilson coefficient has to satisfy 
\begin{equation}
  \sqrt{\left(1 + \frac{\Real \, C_{B\bar{B}}^{\text{NP}}}{C_{B\bar{B}}^{\text{SM}}} \right)^{2} +          \left(\frac{\Imag \, C_{B\bar{B}}^{\text{NP}}}{C_{B\bar{B}}^{\text{SM}}} \right)^{2}} \in [0.777, 0.998] \quad (2\sigma) \, .
  \label{eq:cbb}
\end{equation}
CP-violating effects are further constrained by the CP asymmetry of the golden mode $B_{s} \rightarrow J/\psi \, \phi$ \cite{Amhis:2016xyh},
\begin{equation}
A_{\text{CP}}^{\text{mix}}(B_{s} \rightarrow J/\psi \phi) = \sin(\phi_{\Delta} - 2\beta_{s}) \, = -0.021 \pm 0.031\ ,
\label{eq:acp}
\end{equation}
where $\beta_{s} = 0.01852 \pm 0.00032$\cite{Charles:2004jd}, and penguin contributions are neglected. 
Using Eq.~(\ref{eq:deltadef}), this can be interpreted as an additional constraint on the real and imaginary parts of $C_{B\bar{B}}^{\text{NP}}$ (and in turn, on the real and imaginary parts of the couplings $\lambda_{s}\lambda_{b}^{*}$).

\begin{figure}[!t]
\begin{center}
	\includegraphics[scale=.4]{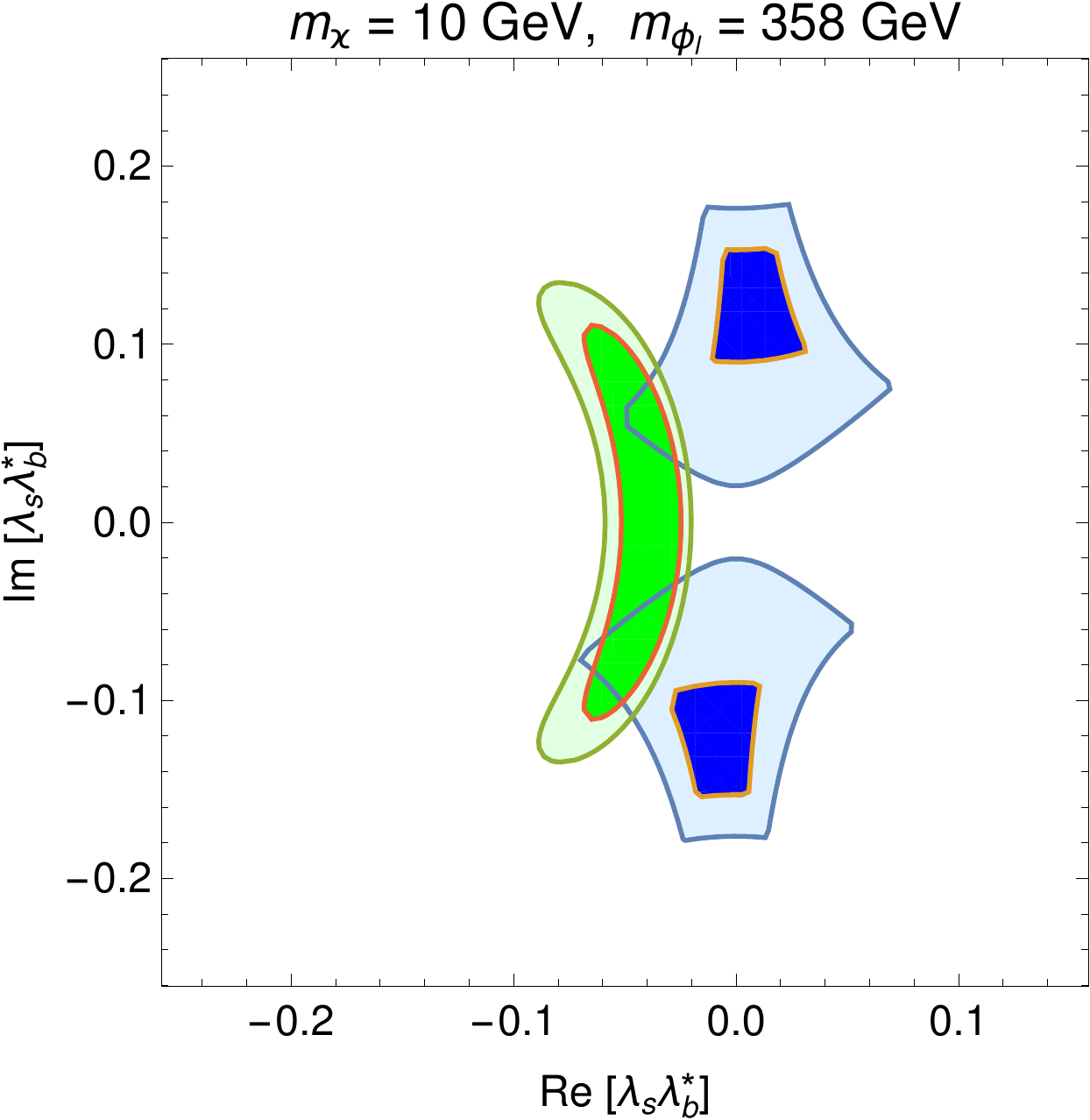}
	\includegraphics[scale=.4]{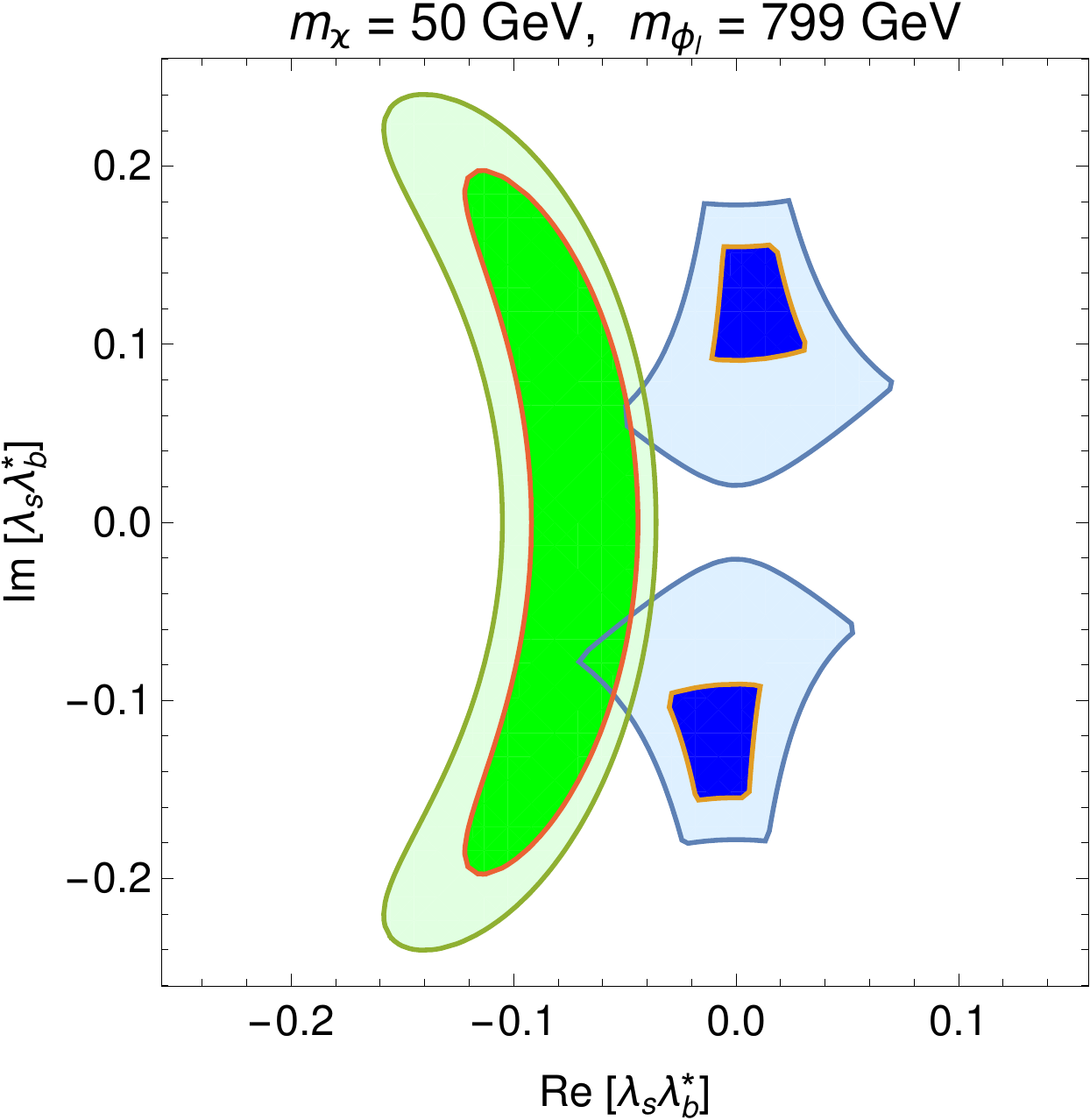}
	\includegraphics[scale=.4]{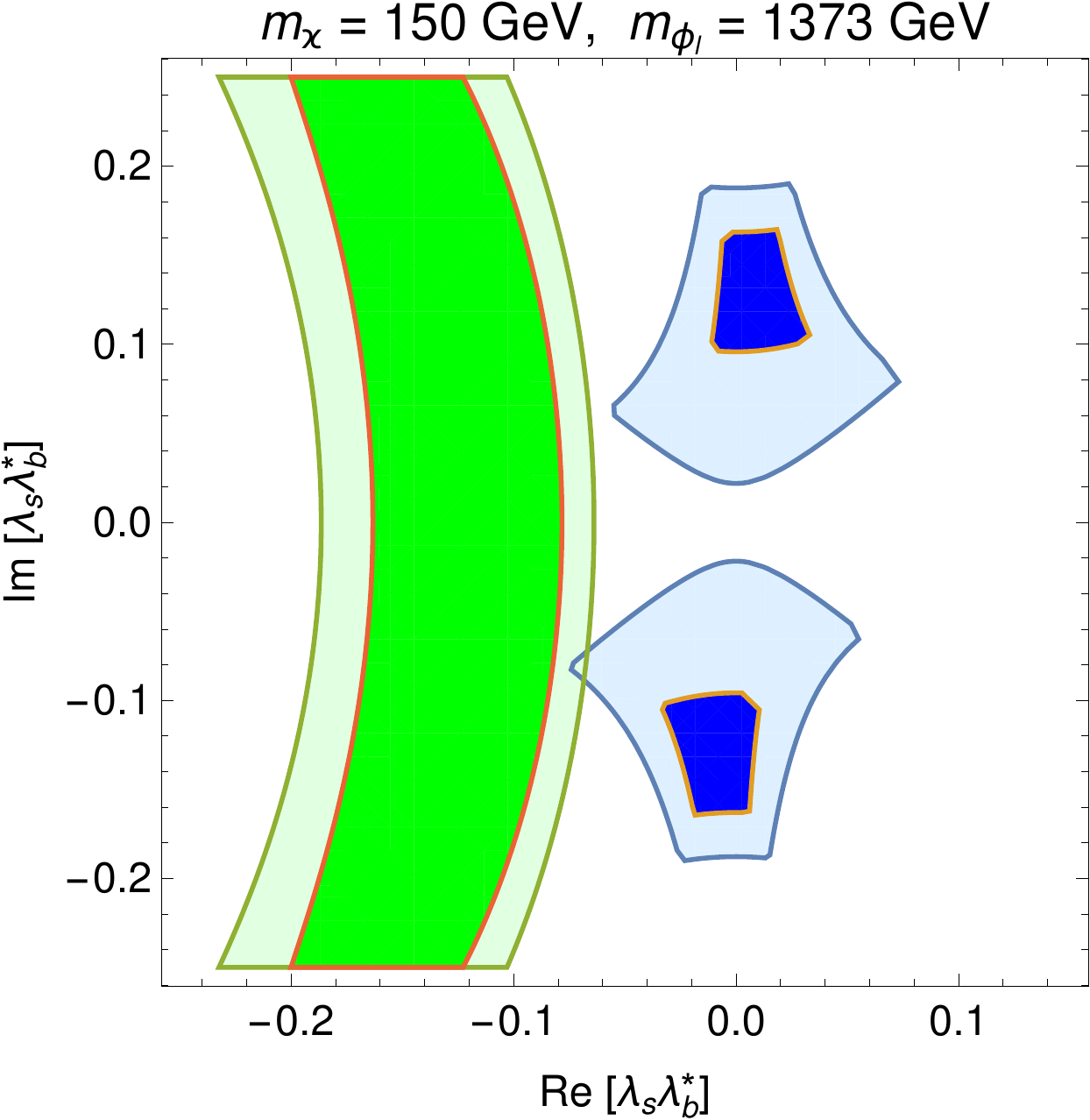}\\ \vspace*{4ex}
	\includegraphics[scale=.4]{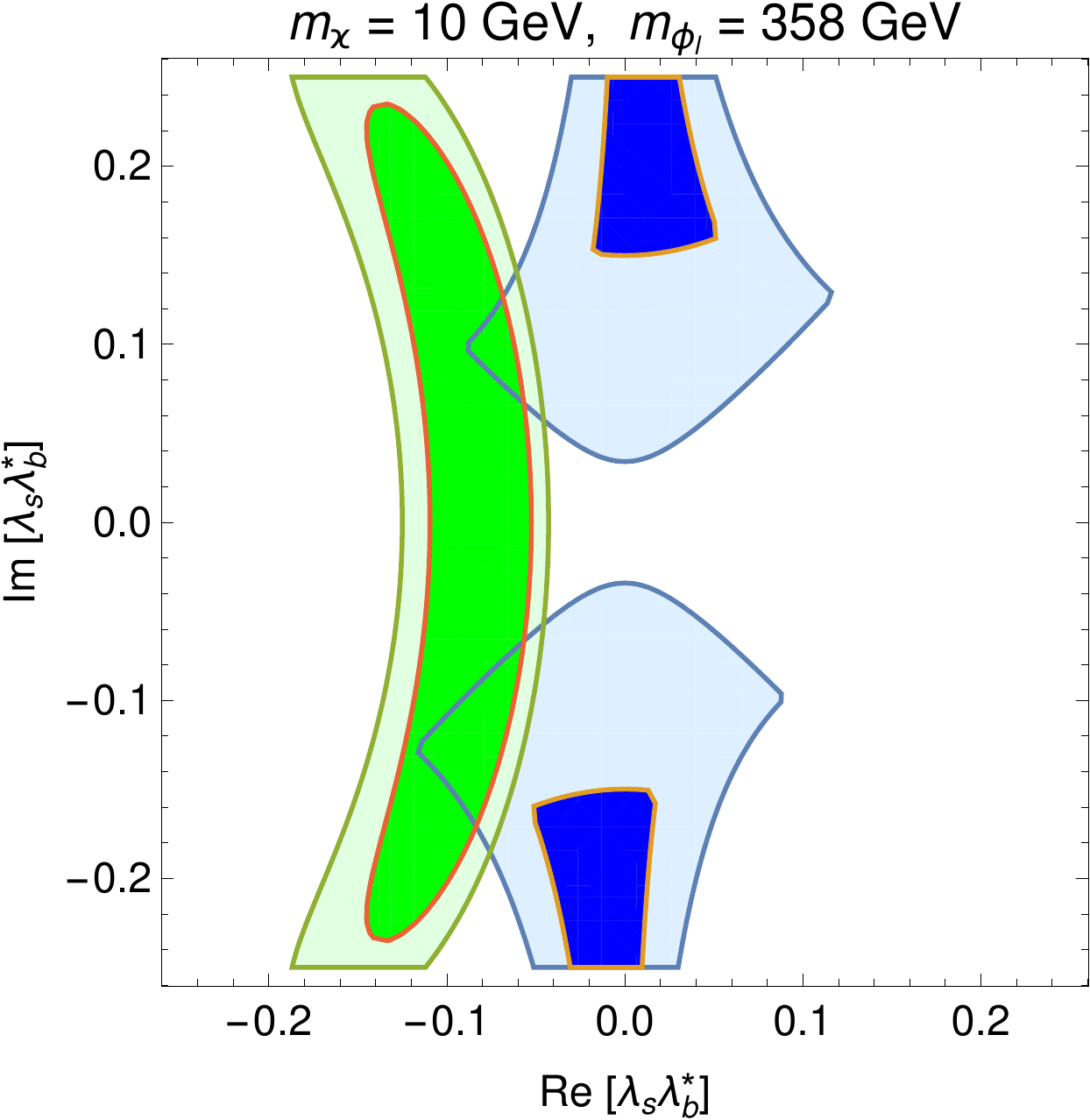}
    \includegraphics[scale=.4]{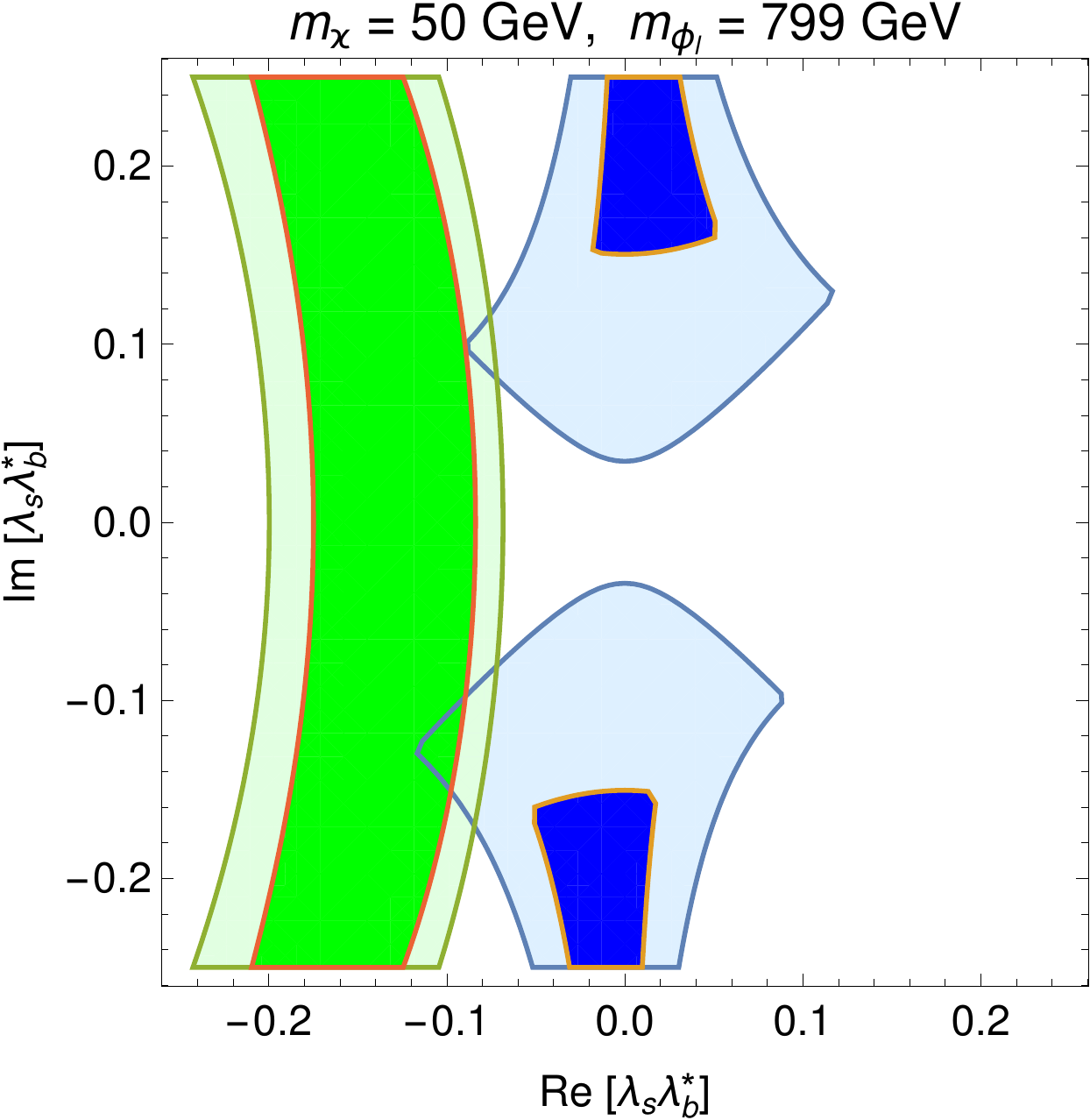}
    \includegraphics[scale=.4]{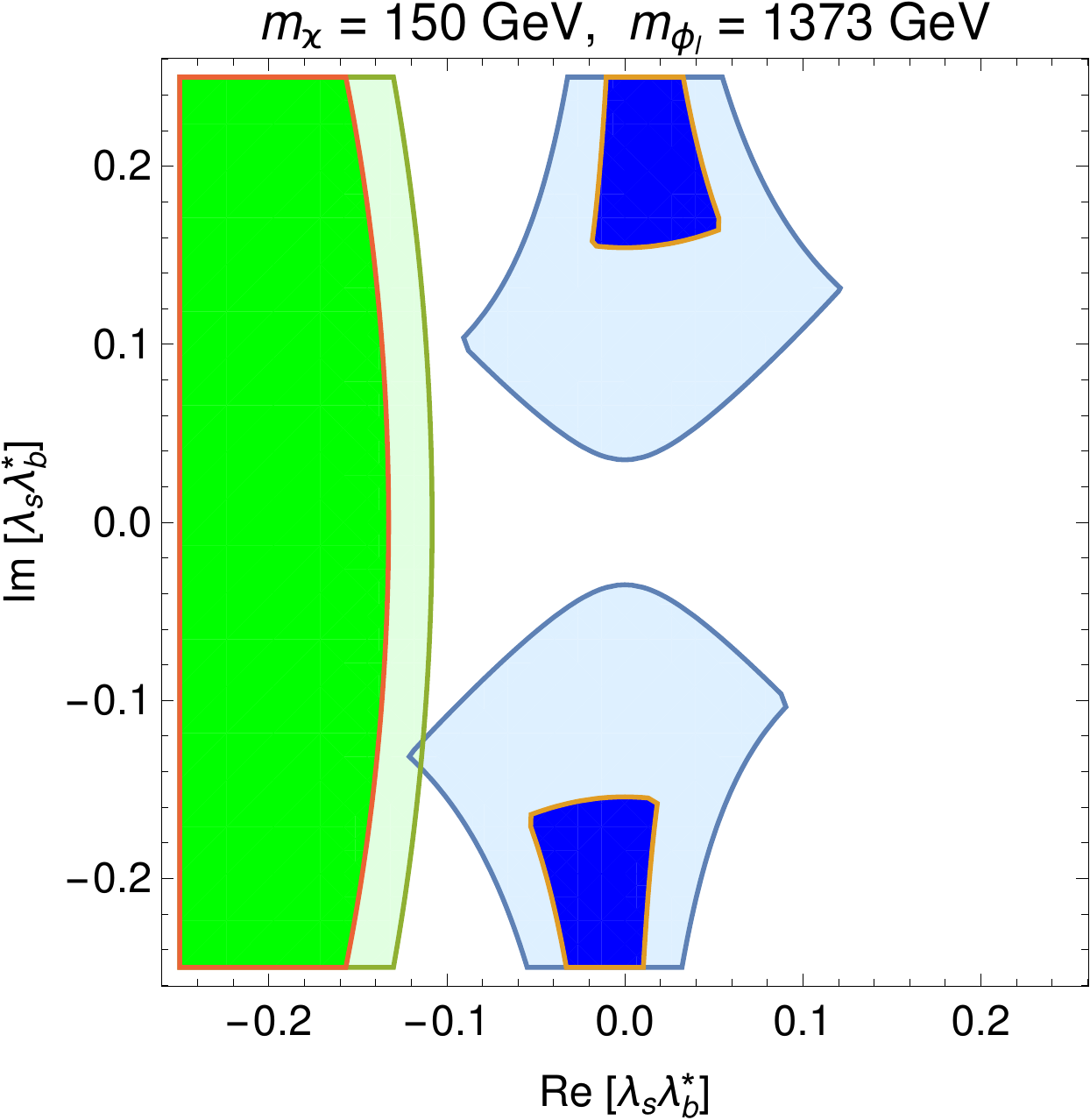}
\end{center}
	\caption{
	 The dark (light) green area is the $1\sigma$ ($2\sigma$) allowed region by $b \rightarrow s \mu^+ \mu^- $  observables in the $(\operatorname{Re}(\lambda_s  \lambda_b^*), \operatorname{Im}(\lambda_s  \lambda_b^*) )$ plane. Dark (light) blue regions correspond to  $1\sigma \; (2\sigma)$  $B_s-$ mixing allowed regions. We take $\lambda_\mu = \sqrt{4\pi} $ and $m_{\phi_q} = 1.5 \, \TeV$ (top row),  $2.5 \, \TeV$ (bottom row). The  specific values of $m_\chi$, $m_{\phi_q}$ are given in the plot and $m_{\phi_l}$ is fixed to reproduce the measured DM relic abundance.
	}
	\label{fig:greensandblues1}
\end{figure}

In Figure~\ref{fig:greensandblues1}, the effect of all of these constraints on the real and imaginary parts of the couplings $\lambda_{s}\lambda_{b}^{*}$ for several benchmark points is shown. Regions that are allowed by $\bsmumu$ observables and $B_{s}-$mixing (given by Eqs.~(\ref{eq:cbb}) and (\ref{eq:acp})) are shaded in green and blue, respectively. For illustrative purposes, the figure shows the constraints for multiple values of DM and mediator masses, while keeping $\lambda_\mu=\sqrt{4\pi}$ fixed. We remind the reader that the mass of $\mphil$ is fixed so as to reproduce the correct relic density using Eq.~(\ref{eq:sigmav}).

As we can observe, in order to simultaneously satisfy both types of constraints, complex couplings are needed ($\Imag(\lambda_{s}\lambda_{b}^{*})\ne 0$). Also, as the mass of the dark matter particle and the mediators increase, both areas are more difficult to reconcile. In practise, this leads to an upper bound on the masses of the exotic new particles. The precise limit depends on the choice of couplings, which we will discuss in Section~\ref{sec:lhc}.

Finally, the new physics couplings to the up-type quarks are generated via CKM rotation,
\begin{eqnarray}
\lambda_{u} &=& V_{us}\lambda_{s} + V_{ub}\lambda_{b} \, , \nonumber\\
\lambda_{c} &=& V_{cs}\lambda_{s} + V_{cb}\lambda_{b} \, .
\label{eq:upcharmCKM}
\end{eqnarray} 
These couplings generate a new physics contribution to $D^{0}-$mixing, and the Wilson coefficient $C_{D\bar{D}}^{\text{NP}}$ is obtained replacing $\lambda_{s}$ and $\lambda_{b}^{*}$ in Eq.~(\ref{eq:wilsonCBB}) by $\lambda_{u}$ and $\lambda_{c}^{*}$, respectively.

In contrast to $B_{s}-$mixing, there is no precise theory determination for the mass difference in the $D^{0}$ system. Therefore, in order to constrain the new physics contribution to $C_{D\bar{D}}$ we use the measured value of the mass difference in $D^{0}-$mixing. The experimental bound on the mixing diagram is given by \cite{Bona:2017kam}
\begin{equation}\label{eq:M12exp}
  |M_{12}|_{D\bar{D}}^{\text{exp}} \in [0.6, 7.5] \times 10^{-3} \, \text{ps}^{-1} \quad (2\sigma) \, ,
\end{equation}
whereas the new physics contribution to $D^{0}-$mixing is described by
\begin{equation}\label{eq:M12lattice}
  |M_{12}|_{D\bar{D}} = \frac{|C_{D\bar{D}}|}{2M_{D^{0}}}\langle{D^{0}}| \pazocal{O} |{\bar{D}^{0}}\rangle \, ,
\end{equation}
where $\pazocal{O}$ is a combination of operators containing all possible SM and new physics contributions to $D^{0}-$mixing. Using the last results from \cite{Bazavov:2017weg} we get the following bound on the Wilson coefficient:
\begin{equation}\label{eq:CDD}
  |C_{D\bar{D}}^{\text{exp}}| \le 5.695 \times 10^{-8} \; \TeV^{-2} \quad (2\sigma) \, .
\end{equation}

Although this model induces new physics contributions to other flavour observables (such as $b \rightarrow s\gamma$, $b \rightarrow s\nu\bar{\nu}$ and effective $Z\mu^{+}\mu^{-}$ and $Zq_{i}q_{j}$ couplings), their size is very small and does not produce significant deviations from current experimental searches.

\subsection{Benchmark scenarios}
\label{sec:benchmark}

All the new physics contributions to the observables described above depend on five independent parameters: the three masses of the new particles, $m_{\chi}$, $m_{\phi_{q}}$ and $m_{\phi_{l}}$, the product of the couplings $\lambda_{s}\lambda_{b}^{*}$ and the absolute value of the coupling $|\lambda_{\mu}|$.

The three masses only enter the Wilson coefficients through the factor $m_{\chi}^{-2}$ and the dimensionless loop functions. In addition, all the Wilson coefficients are proportional to $\lambda_{s}\lambda_{b}^{*}$ or $|\lambda_{\mu}|^{2}$ or both. In order to constrain our model, we consider two scenarios by fixing the value of $|\lambda_{\mu}|$. Then we scan over the mass parameters $m_{\chi}$ and $m_{\phi_{q}}$, with $m_{\phi_{l}}$ fixed by the requirement of reproducing the correct relic abundance, and check all the flavour observables described in Section \ref*{sec:bsmixing}. In this way, for any combination of masses and a fixed value of $|\lambda_{\mu}|$ we get a set of allowed values for $\lambda_{s}\lambda_{b}^{*}$. 
We consider two hierarchies between $|\lambda_{s}|$ and $|\lambda_{b}|$ that lead to different constraints from $D^{0}-$mixing, and, ensuring that $\Imag(\lambda_s\lambda_b^*)\ne0$, we define the following benchmark scenarios:

\begin{minipage}{0.48\linewidth}
    \begin{itemize}
        \item[(A1)] $|\lambda_{\mu}| = 2$, with $\lambda_{b} = \lambda_{s}^*$;
        
        \item[(A2)] $|\lambda_{\mu}| = 2$, with $\lambda_{b} = 4\lambda_{s}^*$;
    \end{itemize}
\end{minipage}    
\begin{minipage}{0.48\linewidth}
    \begin{itemize}
        \item[(B1)]  $|\lambda_{\mu}| = \sqrt{4\pi}$, with $\lambda_{b} = \lambda_{s}^*$;

        \item[(B2)]  $|\lambda_{\mu}| = \sqrt{4\pi}$, with $\lambda_{b} = 4\lambda_{s}^*$,
    \end{itemize}
\end{minipage}

\noindent
where $|\lambda_{\mu}| = \sqrt{4\pi}$ is the perturbative limit. After establishing a hierarchy between $|\lambda_{s}|$ and $|\lambda_{b}|$, we calculate their maximum and minimum allowed values from the corresponding maximum and minimum allowed values of $\lambda_{s}\lambda_{b}^{*}$. Scenarios with $|\lambda_{s}| > |\lambda_{b}|$ are excluded by $D^{0}-$mixing constraints. Likewise, as we will see in Section~\ref{sec:lhc}, smaller values of $\lambda_\mu$ are constrained by LHC bounds.

\section{LHC constraints and prospects for high-luminosity}
\label{sec:lhc}

In this section, we study the experimental signatures that this model would produce at the LHC. DM search strategies in both ATLAS and CMS involve analysing final states containing jets and leptons produced in association with a DM particle, identified from missing transverse energy. In this model, direct production of the coloured and leptonic scalar doublets $\phi_q$ and $\phi_l$, respectively, typically leads to such final states.

\begin{figure}[t!]
\centering
\renewcommand{\thesubfigure}{a1} 
\begin{subfigure}[t]{.32\textwidth}
  \centering
  \includegraphics[width=\linewidth]{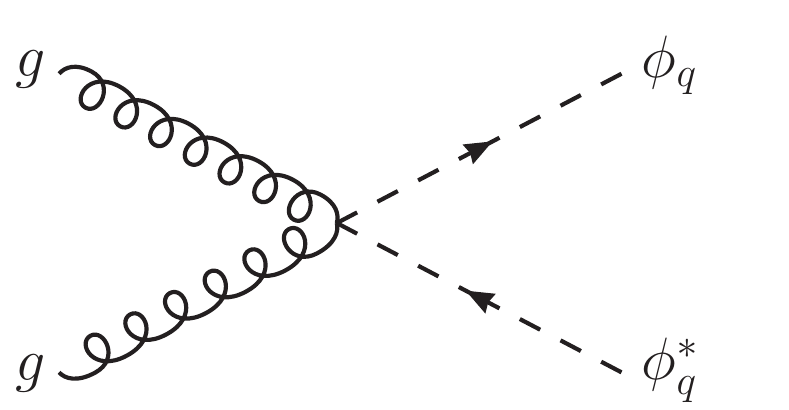}
  \caption{}
\end{subfigure}
\hspace{-10pt}    
\renewcommand{\thesubfigure}{a2}
\begin{subfigure}[t]{.32\textwidth}
  \centering
  \includegraphics[width=\linewidth]{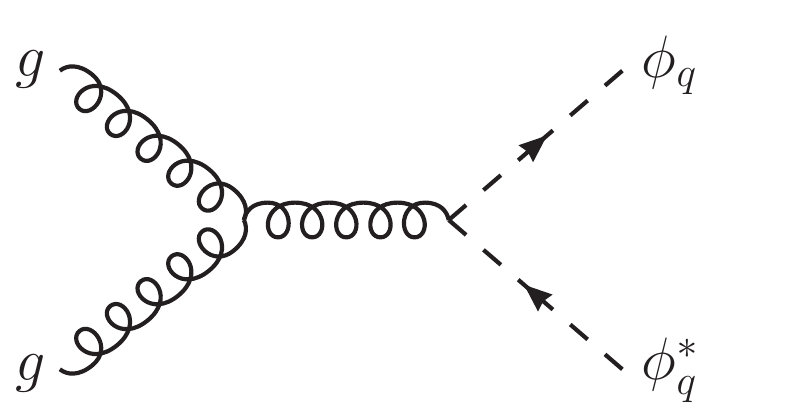}
  \caption{}
\end{subfigure}
\hspace{-10pt}
\renewcommand{\thesubfigure}{a3}
\begin{subfigure}[t]{.32\textwidth}
  \centering
  \includegraphics[width=\linewidth]{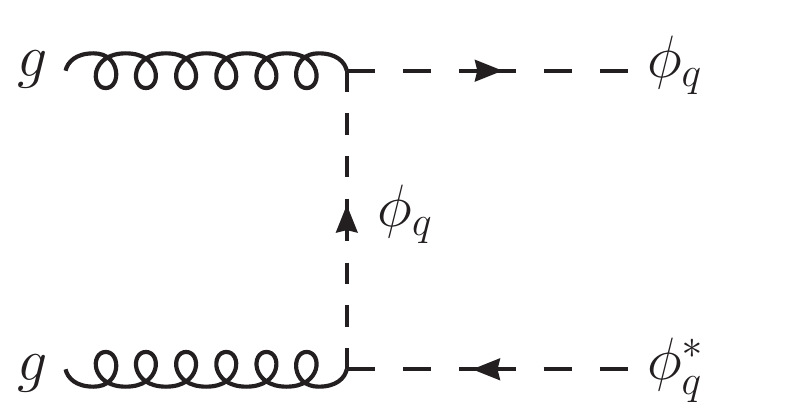}
  \caption{}
\end{subfigure}

\vspace{0pt}
\renewcommand{\thesubfigure}{a4}
\begin{subfigure}[t]{0.32\textwidth}
  \centering
  \includegraphics[width=\linewidth]{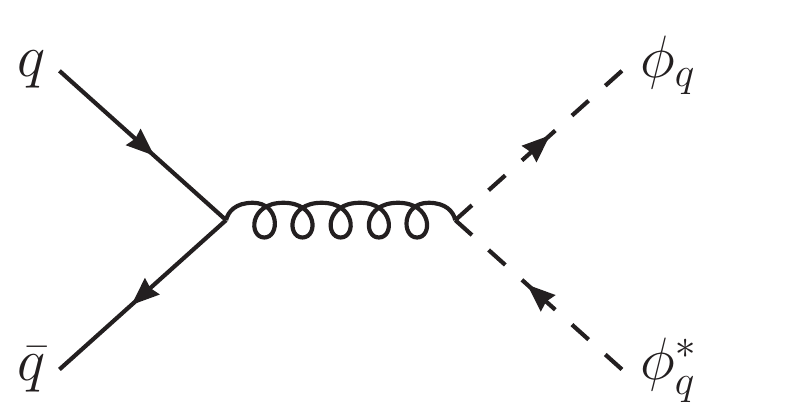}
  \caption{}
\end{subfigure}
\hspace{-10pt}
\renewcommand{\thesubfigure}{b}
\begin{subfigure}[t]{0.32\textwidth}
  \centering
  \includegraphics[width=\linewidth]{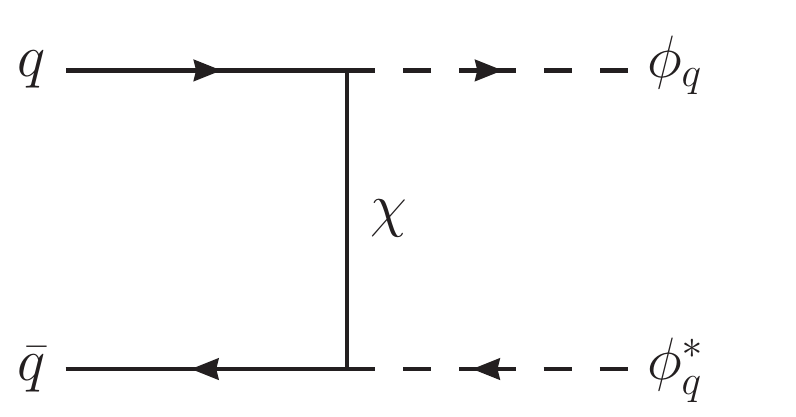}
  \caption{}
\end{subfigure}
\hspace{-10pt}
\renewcommand{\thesubfigure}{c1}
\begin{subfigure}[t]{0.32\textwidth}
  \centering
  \includegraphics[width=\linewidth]{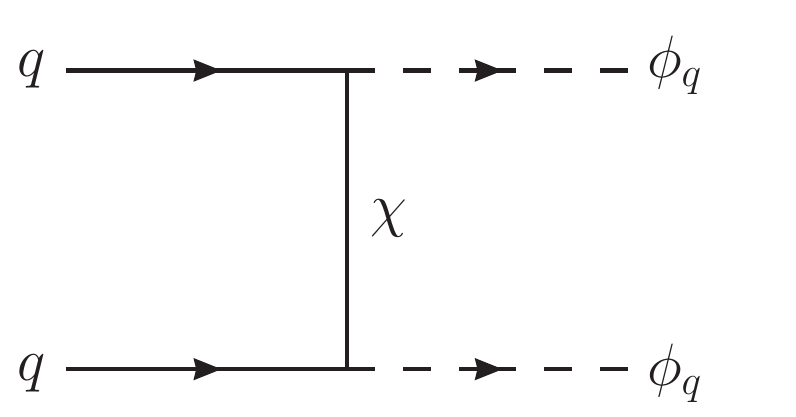}
  \caption{}
\end{subfigure}

\vspace{0pt}
\renewcommand{\thesubfigure}{c2}
\begin{subfigure}[t]{0.32\textwidth}
  \centering
  \includegraphics[width=\linewidth]{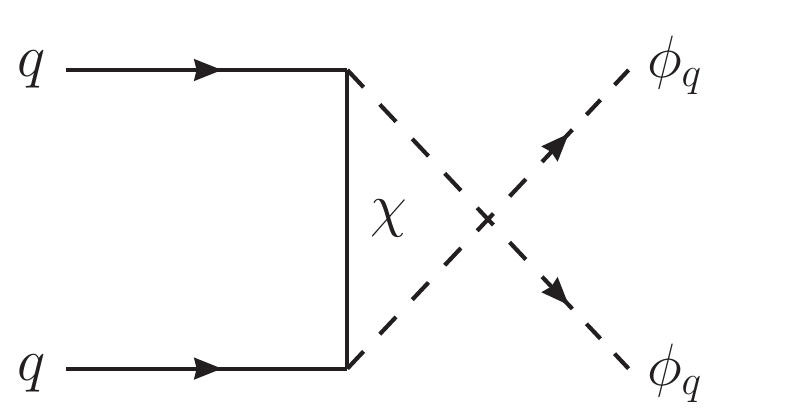}
  \caption{}
\end{subfigure}
\hspace{-10pt}
\renewcommand{\thesubfigure}{c3}
\begin{subfigure}[t]{0.32\textwidth}
  \centering
  \includegraphics[width=\linewidth]{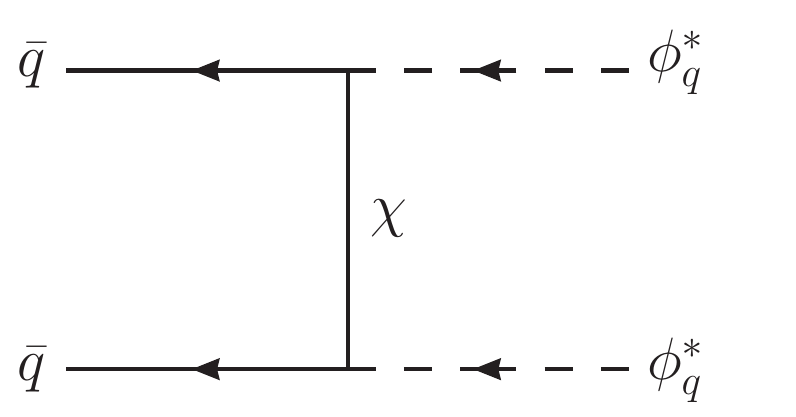}
  \caption{}
\end{subfigure}
\hspace{-10pt}
\renewcommand{\thesubfigure}{c4}
\begin{subfigure}[t]{0.32\textwidth}
  \centering
  \includegraphics[width=\linewidth]{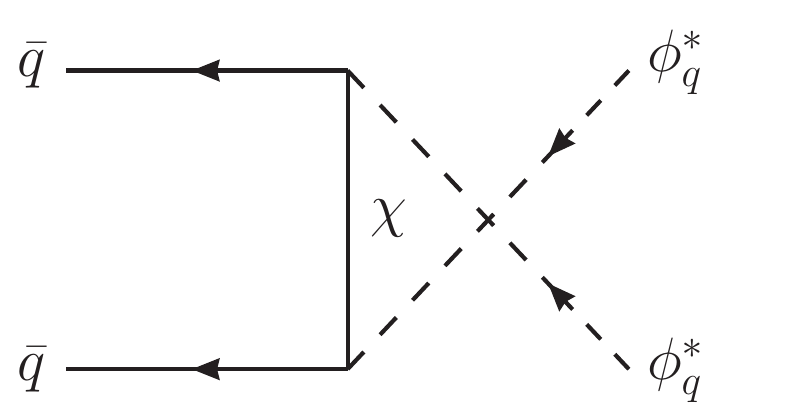}
  \caption{}
\end{subfigure}
\caption{Diagrams for the pair production of the coloured scalar mediator, $\phi_q$,  leading to $\text{dijet} + \MET$ signatures in the final state.  Diagrams (a1)--(a4) are generated by purely QCD interactions, and diagrams (b), (c1)--(c4) are generated by DM t-channel exchange.}\label{fig:prod-colour}
\end{figure}

Let us first consider production processes that involve the coloured scalar, $\phi_q$. In this case, our model could lead to visible signals in final states with both $\text{monojet} \, / \, \text{dijet} + \MET$ signatures. When the new physics coupling $\lambda_q$ is smaller than the strong interaction coupling, $\alpha_{\text{QCD}}$, pure QCD processes constitute the main contribution to the cross section \cite{An:2013xka}. In this model, this implies that QCD diagrams dominate over those with new physics couplings. As a consequence, monojet searches for this model are less effective than dijet searches and we will concentrate on the latter. The $\text{dijet} + \MET$ processes are shown in Figure~\ref{fig:prod-colour}, where diagrams (a) correspond to the QCD contributions, and diagrams (b) and (c) involve new physics couplings. The main production channel is the pair production of the coloured scalar particles, that subsequently decays into a DM particle and a quark,
\begin{equation}
pp \to \phi_q\phi_q^* \, / \, \phi_q\phi_q \, / \, \phi_q^*\phi_q^* \to q q + \MET \; .
\end{equation}
In addition, the scalar doublet $\phi_q$ has the same quantum numbers as squarks in supersymmetric (SUSY) models. Therefore, the kinematics in its production and decay in diagrams (a) of Figure~\ref{fig:prod-colour} mimic those of squarks in SUSY models with decoupled gluinos. As a consequence, limits from ATLAS and CMS squark searches can be used to constrain the model.

\begin{figure}[t!]
\centering
\renewcommand{\thesubfigure}{d1} 
\begin{subfigure}[t]{.32\textwidth}
  \centering
  \includegraphics[width=\linewidth]{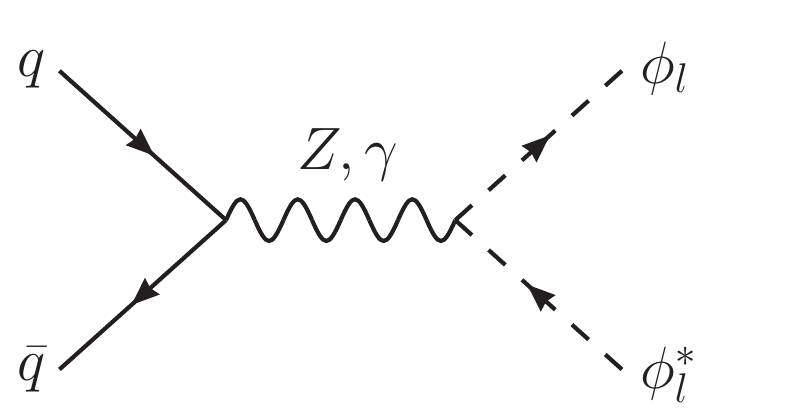}
  \caption{}
\end{subfigure}
\hspace{0pt}    
\renewcommand{\thesubfigure}{d2}
\begin{subfigure}[t]{.32\textwidth}
  \centering
  \includegraphics[width=\linewidth]{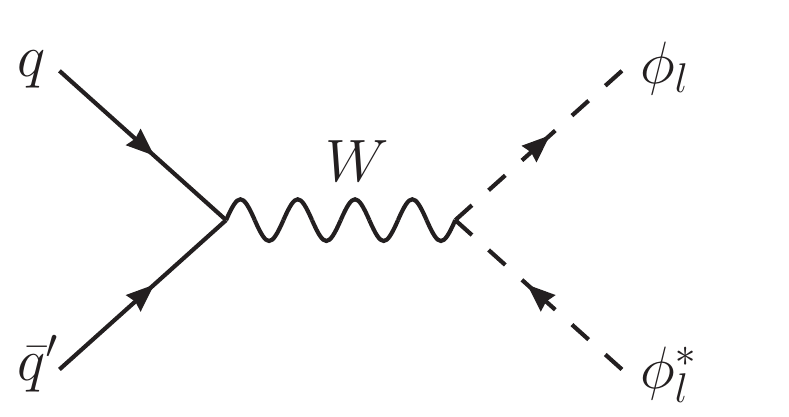}
  \caption{}
\end{subfigure}
\caption{Diagrams for the pair production of the leptonic scalar mediator, $\phi_l$, leading to $\mu \mu / \mu \nu + \MET$ signatures in the final state.}\label{fig:prod-ew}
\end{figure}

One can also consider the pair production of the leptonic scalar, $\phi_l$. In this case, the production process is mediated by $W$ or $Z$ bosons and involves the electroweak coupling, as shown in Figure~\ref{fig:prod-ew}. The decays of $\phi_l$ lead to clean final states with one or two leptons and missing energy. Although flavour constraints require $\lambda_\mu \gg \lambda_q$, the cross section of this process is smaller than the production of the coloured mediator for similar mediator masses. However, since $m_{\phi_l}$ is fixed for every value of $m_{\chi}$ to reproduce the correct relic abundance, there are regions of the parameter space where both searches are complementary. We will here consider the  process
\begin{equation}
pp \to \phi_l\phi_l^* \to \mu \mu / \mu \nu + \MET \; ,
\end{equation}
where the dimuon channel leads to the strongest constraints. As in the previous case, we can exploit the analogy between $\phi_l$ and sleptons to use the limits from slepton searches to constrain this model.

\subsection{Simulation details}

We have implemented this model in \texttt{Feynrules~2.3}\cite{Alloul:2013bka}. The calculation of the matrix elements and the event generation is done using \texttt{MadGraph5\_aMC@NLO~2.6.3}\cite{Alwall:2014hca}. Production and decay of the new particles are considered independently using the narrow width approximation, as implemented in \texttt{MadSpin}\cite{Artoisenet:2012st}, which further accounts for spin correlations in decay chains\footnote{The narrow width approximation is not valid in benchmark points B1 and B2, for which we have taken interference effects into account. }. We then use \texttt{Pythia 8.235}\cite{Sjostrand:2014zea} to shower the parton-level events and we pass the output to \texttt{CheckMATE 2.0.26} \cite{Dercks:2016npn}, which compares the expected signal with supersymmetric searches at the LHC and derives an exclusion limit. As we have explained above, we can apply squark and slepton searches to constrain the coloured and leptonic mediator, respectively.

In order to describe initial and final state radiation and reproduce the correct jet structure precisely, we consider leading order (LO) production with parton shower matching and multijet merging when needed. The LO multijet merging techniques describe how parton shower emissions can be combined with full matrix element calculations to achieve a better accuracy in the description of the radiation spectrum. Using this technique, every jet is classified according to its $p_T$ and then compared to a hardness scale $Q_{\text cut}$. In this way, emissions above the hardness scale $Q_{\text cut}$ are described at LO accuracy using the corresponding matrix element calculation for an extra hard, wide-angle QCD emission in the final state, while emissions below this scale are defined as soft or collinear jets and the all-orders resummation description from the parton shower is preserved. Note that even though $\pazocal{O}(\alpha_s)$ corrections are included using this procedure, the calculation remains formally $\text{LO}+\text{LL}$ accurate after parton shower due to missing virtual corrections.

After hadronization, the showered events and the production cross sections are passed to \texttt{CheckMATE}. Each model point is tested against all the implemented experimental analyses to determine the optimal signal region. For this signal region, \texttt{CheckMATE} compares the simulated signal with the actual experimental observation and determines whether the model point is excluded at the $90\%$ confidence level.

\subsection{Results}

\begin{figure}[t!]
\centering
  \includegraphics[width=0.49\linewidth]{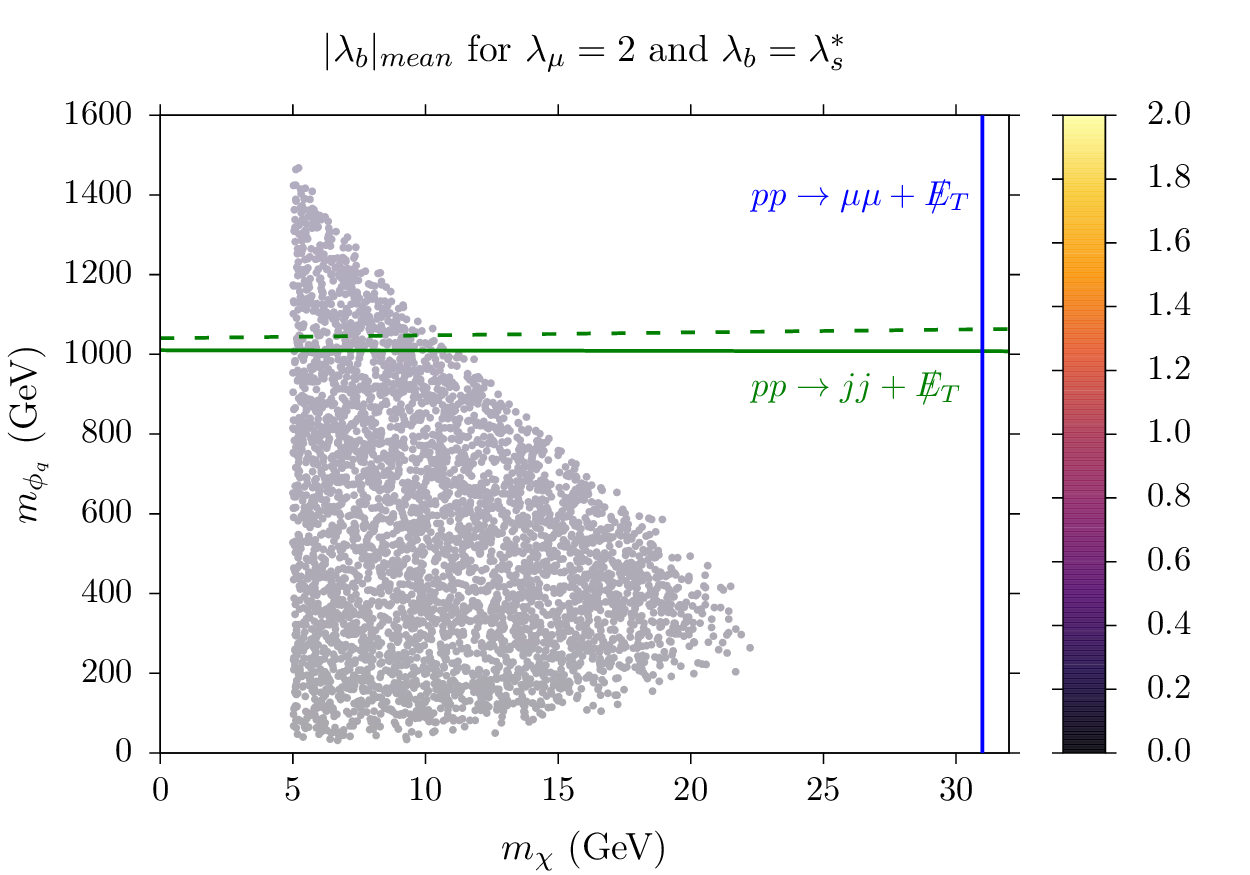}
  \includegraphics[width=0.49\linewidth]{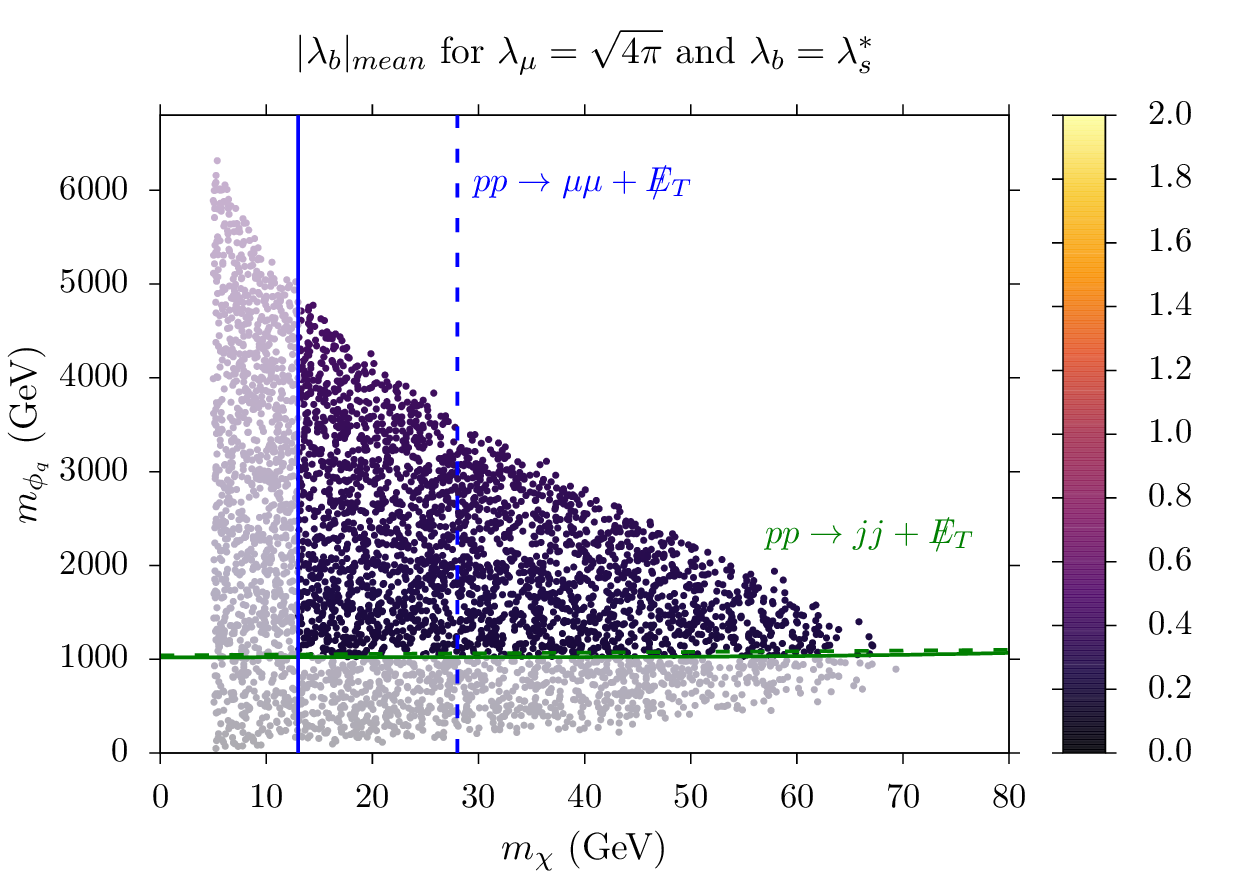}\\[2ex]
  \includegraphics[width=0.49\linewidth]{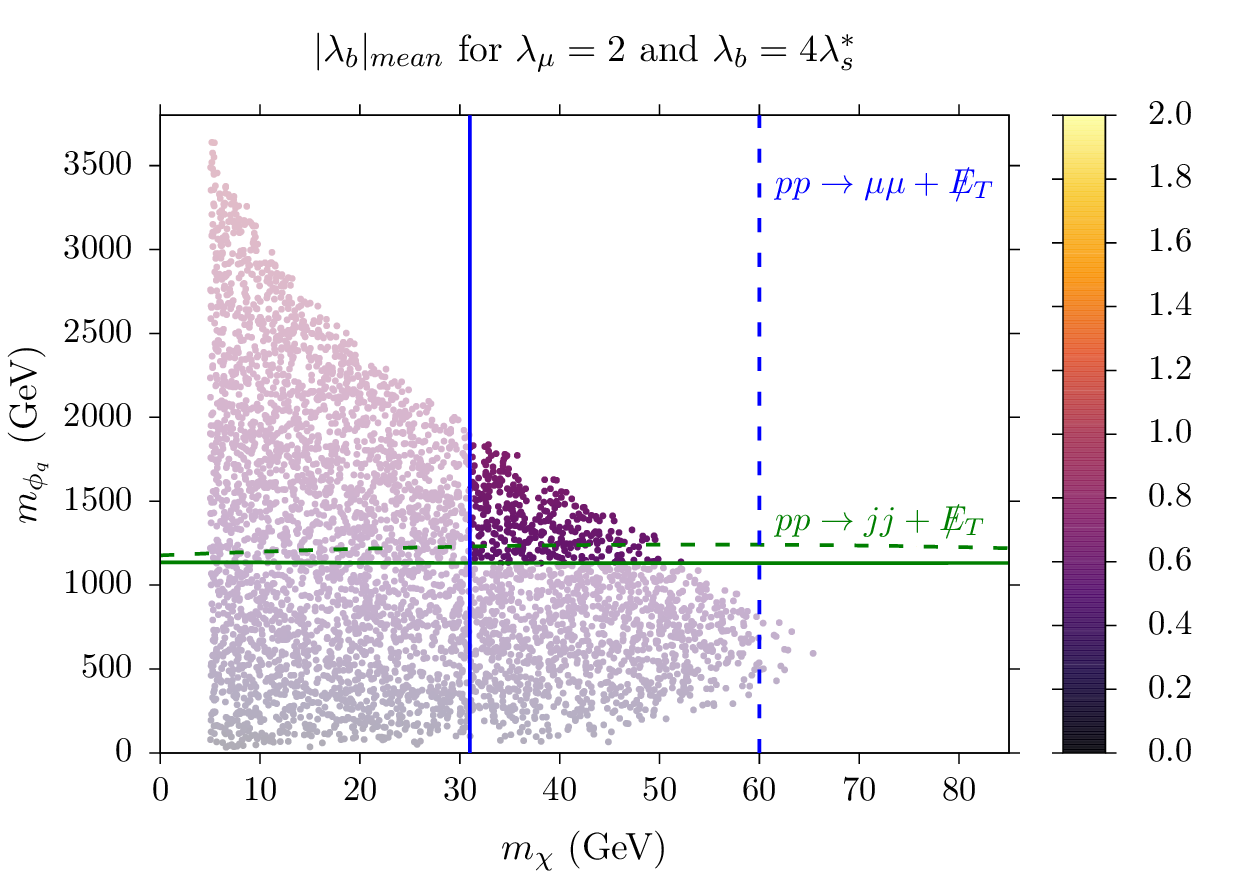}
  \includegraphics[width=0.49\linewidth]{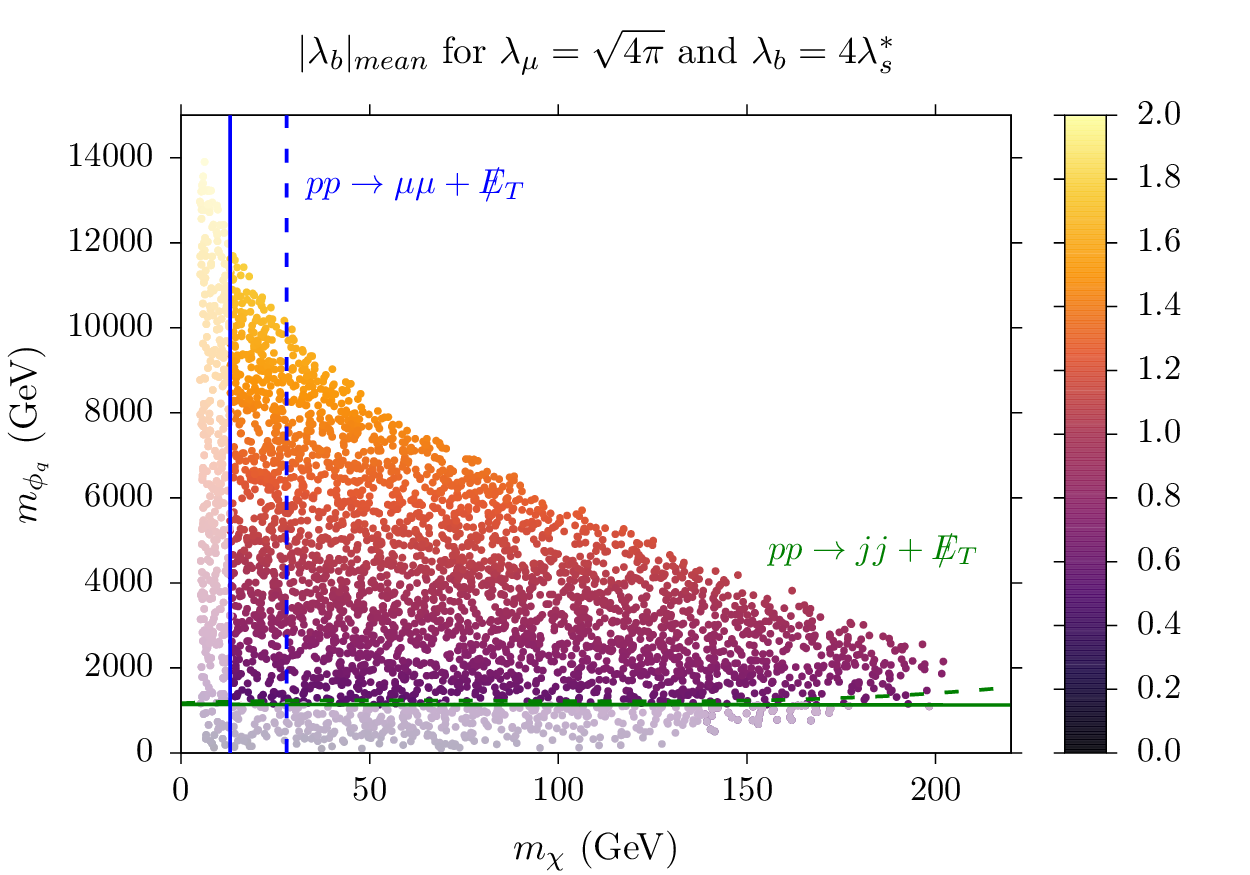}
\caption{LHC limits from the $pp \to jj + \MET$ (green) and $pp \to \mu\mu + \MET$ (blue) searches. On the left (right), results for the scenario with $|\lambda_\mu| = 2$ ($|\lambda_\mu| = \sqrt{4\pi}$) are presented. The figures in the upper panel correspond to $\lambda_b = \lambda_s^*$, while the bottom panel shows limits for $\lambda_b = 4\lambda_s^*$. The colour code represents the average value of the coupling $|\lambda_b|$ in the region allowed by flavour constraints, as defined in the text. Solid lines represent the current exclusion limits, whereas dashed ones correspond to the projected reach of the LHC High luminosity phase. }\label{fig:lhc}
\end{figure}

Constraints from LHC searches for the four benchmark points defined in Section~\ref{sec:benchmark} are presented in Figure~\ref{fig:lhc} on the $(\mchi,\,\mphiq)$ plane, for all the points that satisfy the flavour constraints of Section~\ref{sec:bsmixing} and that reproduce the correct DM relic abundance. This figure shows the complementarity between the experimental limits obtained from the $pp \to jj + \MET$ and $pp \to \mu\mu + \MET$ searches. The experimental results used in our analysis are summarised in Table~\ref{tab:lhc_searches}. The colour code represents the average value of the coupling $|\lambda_b|$ in the region allowed by flavour constraints, defined as $|\lambda_b|_{\rm mean} = (|\lambda_b|_{\rm max} + |\lambda_b|_{\rm min}) / 2$, where $|\lambda_b|_{\rm max}$ and $|\lambda_b|_{\rm min}$ are the maximum and minimum allowed values respectively. The variation of our results when choosing either the minimum or maximum value for $|\lambda_b|$ has been checked and is insignificant.

Regarding the $pp \to jj + \MET$ search, the limits in every scenario show that for the lightest DM mass, coloured mediators with masses below $\sim$1~TeV are excluded.  Even though heavier DM produces larger amounts of missing energy in final states, the cross section decreases rapidly with the $\mchi$, leading to similar exclusion limits. It is interesting to note that exclusion limits are slightly stronger for the scenarios with $|\lambda_{b,t}| > |\lambda_{s,c}|$, where mediators with masses below $\sim$1.1~TeV are excluded. The reason for this is that final states with either top or bottom quarks are more sensitive to some experimental searches. The most stringent experimental search involves final states with at least two ($b \bar b$ production) or four ($t \bar t$ production) jets or exactly two leptons and missing energy \cite{Aaboud:2017rzf}. In particular, the most sensitive signal region is optimised to detect events featuring a DM particle produced in association with a $t \bar t$ pair, which decays fully hadronically.

\vspace{10pt}
\begin{table}[t!]
\begin{center}
\begin{tabular}{|c|l|l|c|}
\hline
$\sqrt{s}$ & Search & Final state & $\pazocal{L}$ [fb$^{-1}$] \\
\hline
\hline
13 TeV & 1710.11412 \cite{Aaboud:2017rzf} & $t\bar t \, / \, b\bar b + \MET$ & 36.1 \\
 & ATLAS-CONF-2017-039 \cite{ATLAS:2017uun} & $2l \, / \, 3l + \MET$ & 36.1 \\
\hline
\hline
14 TeV & atlas\_phys\_pub\_013\_011 \cite{ATL-PHYS-PUB-2013-011} & $t\bar t + \MET$ & 3000 \\
& atlas\_phys\_2014\_010\_hl\_3l \cite{ATL-PHYS-PUB-2014-010} & $2l \, / \, 3l + \MET$ & 3000 \\
\hline
\end{tabular}
\caption{List of experimental searches sensitive to our model, where $l$ denotes electron and muon. The third column describes the final state targeted by the analysis and the last column displays the total integrated luminosity. }
\label{tab:lhc_searches} 
\end{center}
\end{table}
\vspace{-15pt}

Regarding the $pp \to \mu\mu + \MET$ search, the limits show that models with dark matter masses below approximately $ 30 \; \GeV$ are ruled out for $|\lambda_\mu| = 2$, with the exclusion limit going down to $\sim 13 \; \GeV$ for $|\lambda_\mu| = \sqrt{4\pi}$. This corresponds to mediator masses below $360 \; \GeV$ for $|\lambda_\mu| = 2$ and $410 \; \GeV$ for $|\lambda_\mu| = \sqrt{4\pi}$. The $pp \to \mu\mu + \MET$ cross section mainly depends on $m_{\phi_l}$, so the limits on $m_\chi$  can be understood through its relation with  $m_{\phi_l}$ given by the DM relic condition  (\ref{eq:sigmav}) for a particular value of $\lambda_\mu$. The most stringent search involves final states with $2l + 0 j$, $2l$ and at least $2$ jets, or $3l$ and missing energy \cite{ATLAS:2017uun}. In particular, the most sensitive signal region is characterised by $2l + 0j$ and a dilepton invariant mass $m_{ll} > 300 \; \GeV$, and it is optimised to target slepton pair production.

The most remarkable result is that LHC limits completely exclude the scenario with $|\lambda_\mu| = 2$ and $\lambda_b = \lambda_s^*$, as well as a sizeable region of the scenario with $\lambda_b = 4\lambda_s^*$ for the same $|\lambda_\mu|$. These constraints become weaker for larger values of $|\lambda_\mu|$ and, for the scenarios with $|\lambda_\mu| = \sqrt{4\pi}$, most of the parameter space is allowed. It is crucial to note that the limits coming from final states with jets and leptons are complementary to each other. While the former exclude regions of the parameter space with large $m_\chi$ and small $m_{\phi_q}$, the latter rule out models with very heavy mediator masses $m_{\phi_q}$ and light dark matter. Importantly, these limits are also complementary to the ones coming from direct detection, where dark matter masses below $12 \; \GeV$ lie below the neutrino floor. Therefore, it is fundamental to consider both approaches to explore the model.

It is worth mentioning that the small couplings required by flavour constraints lead to decay widths slightly below the QCD scale for $\mphiq \lesssim 370 \; \GeV$. Strictly speaking, this means that the computation of the decay width cannot be handled perturbatively and that the new particle $\phi_q$ may hadronize into bound states with SM quarks, analogous to R-hadrons \cite{FARRAR1978575}, before decaying. However, the typical width involved is $\Gamma_{\phi_q} \sim \pazocal{O}(10^{-2})-\pazocal{O}(10^{-3}) \; \GeV$, which means lifetimes of the order $\tau \sim 10^{-22} \; \text{s}$, so any potential bound state would decay promptly in the detector. This region of the parameter space is excluded by ATLAS and CMS R-hadron searches \cite{Aad:2013gva,CMS-PAS-EXO-16-004}.

We have also studied the limits that could be obtained with $3000 \; \text{fb}^{-1}$ of 14~TeV data once the LHC High Luminosity phase \cite{Apollinari:2017cqg} is completed. As we can observe in the plots, the main gain would come from the leptonic channels, which would allow to test a considerable amount of the model's parameter space. In particular, scenarios with $|\lambda_\mu| < 2$ would be completely excluded. The experimental searches giving the strongest exclusion limits target the same final states and are shown in the low panel of Table \ref{tab:lhc_searches}.

\section{Direct DM detection prospects}
\label{sec:direct}

Finally, in this section we discuss whether our model is expected to produce an observable response in direct detection experiments. We have calculated this response, by matching the model parameters to effective DM-nucleon interaction terms,
\begin{equation}
{\cal L}_{\text{int}}=\sum_{N} \sum_i c_i^{N}\mathcal{O}_i \overline{\chi} \chi \overline{N} N\ ,
\label{eq:eft}
\end{equation}
where $N$ is the corresponding nucleon, and $\mathcal{O}_i$ is the set of non-relativistic operators \cite{Fan:2010gt,Fitzpatrick:2012ix}. The values for the coefficients $c_i^N$ can be derived as the non-relativistic limit of the original interaction Lagrangian, and the differential rate can be computed using the corresponding nuclear form factors from Refs.~\cite{Fitzpatrick:2012ix,Anand:2013yka}, and for a given choice of the DM halo properties. We  have adopted the so-called standard halo model~\cite{Drukier:1986tm} with local DM density $\rho_{\chi}=0.4\ \textrm{GeV/cm}^3$, a central velocity of $v_0=220\ \textrm{km s}^{-1}$, and a escape speed of $v_{\textrm{esc}}=544\ \textrm{km s}^{-1}$ to calculate the number of expected recoils in a specific experiment.

The leading tree-level DM-quark interactions are given by scalar ($\overline{\chi}\chi\overline{\psi}\psi$) and vector ($\overline{\chi}\gamma^{\mu}\chi\overline{\psi}\gamma_{\mu}\psi$) type interactions. The latter is the leading contribution to $\mathcal{O}_1$ for Dirac DM \cite{Ibarra:2015fqa}, but it vanishes in the case of Majorana DM. For scalar type interactions Majorana DM does not in general vanish, but with our models chiral structure, it does.  With sub-dominant couplings to the first generation of quarks, and given that $\mphiq>\mphil$, one-loop contributions to the DM-nucleon scattering cross sections will generally be larger than the tree level process. The loop contributions for a generic fermionic DM that involve the exchange of a photon can be classified as electric and magnetic dipoles ($\overline{\chi}i\sigma^{\mu\nu}\gamma^5\chi F_{\mu\nu}$ and $\overline{\chi}\sigma^{\mu\nu}\chi F_{\mu\nu}$, respectively), anapole ($\overline{\chi}\gamma^{\mu}\gamma^5\chi\partial^{\nu}F_{\mu\nu}$), and charge radius ($\overline{\chi}\gamma^{\mu}\chi\partial^{\nu}F_{\mu\nu}$). However, in the particular case of Majorana DM considered in this work, the magnetic dipole and charge radius effective couplings are forbidden by charge conjugation symmetry. Thus, the dominant one-loop interaction is the anapole moment \cite{Ho:2012bg}. When taking the non-relativistic limit,  the anapole moment gives contributions to the $\mathcal{O}_8$ and $\mathcal{O}_9$ operators \cite{Gresham:2014vja,DelNobile:2018dfg}, which are velocity and momentum dependent. In terms of the fundamental parameters of the model, the corresponding couplings read
\begin{eqnarray}
    c_8&=& 2 e \mathcal{A} Q_N\nonumber \, ,\\
    c_9&=& - e \mathcal{A} g_N \, ,
\end{eqnarray}
where $e$ is the electron charge, $Q_N$ is the nucleon charge, and $g_N$ are the nucleon g-factors ($g_p=5.59$ and $g_n=3.83$). The  effective coupling to the anapole interaction term, $\mathcal{A}$, reads \cite{Kopp:2014tsa}
\begin{eqnarray}
    \mathcal{A} = -\frac{e\,|\lambda_\mu|^2}{96\pi^2\mchi^2}
    \left[
    \frac32 \log\frac{\mu}{\epsilon} 
    - \frac{1+3\mu-3\epsilon}{\sqrt{(\mu-1-\epsilon)^2-4\epsilon}}
    \arctanh{\left(\frac{\sqrt{(\mu-1-\epsilon)^2-4\epsilon}}{\mu-1+\epsilon}\right)}
    \right]\ ,
\end{eqnarray}
with $\mu\equiv \mphil^{2}/\mchi^2$ and $\epsilon\equiv m_l^2/\mchi^2$. The nuclear responses to the $\mathcal{O}_8$ and $\mathcal{O}_9$ operators are markedly weaker than that of $\mathcal{O}_1$, which implies that, in general, the scattering cross section is very small and beyond current experimental limits. Furthermore, because our DM particle interacts with the quark sector, it is not a priori clear that the spin-independent $\mathcal{O}_1$ and spin-dependent $\mathcal{O}_4$ arising from the so-called twist-2 operator \cite{Hisano:2010ct,Hisano:2015bma,jubbthesis} and the axial vector operator respectively are still negligible.

\begin{figure}[t!]
    \centering
    \hspace*{-0.5cm}\includegraphics[width=0.53\textwidth]{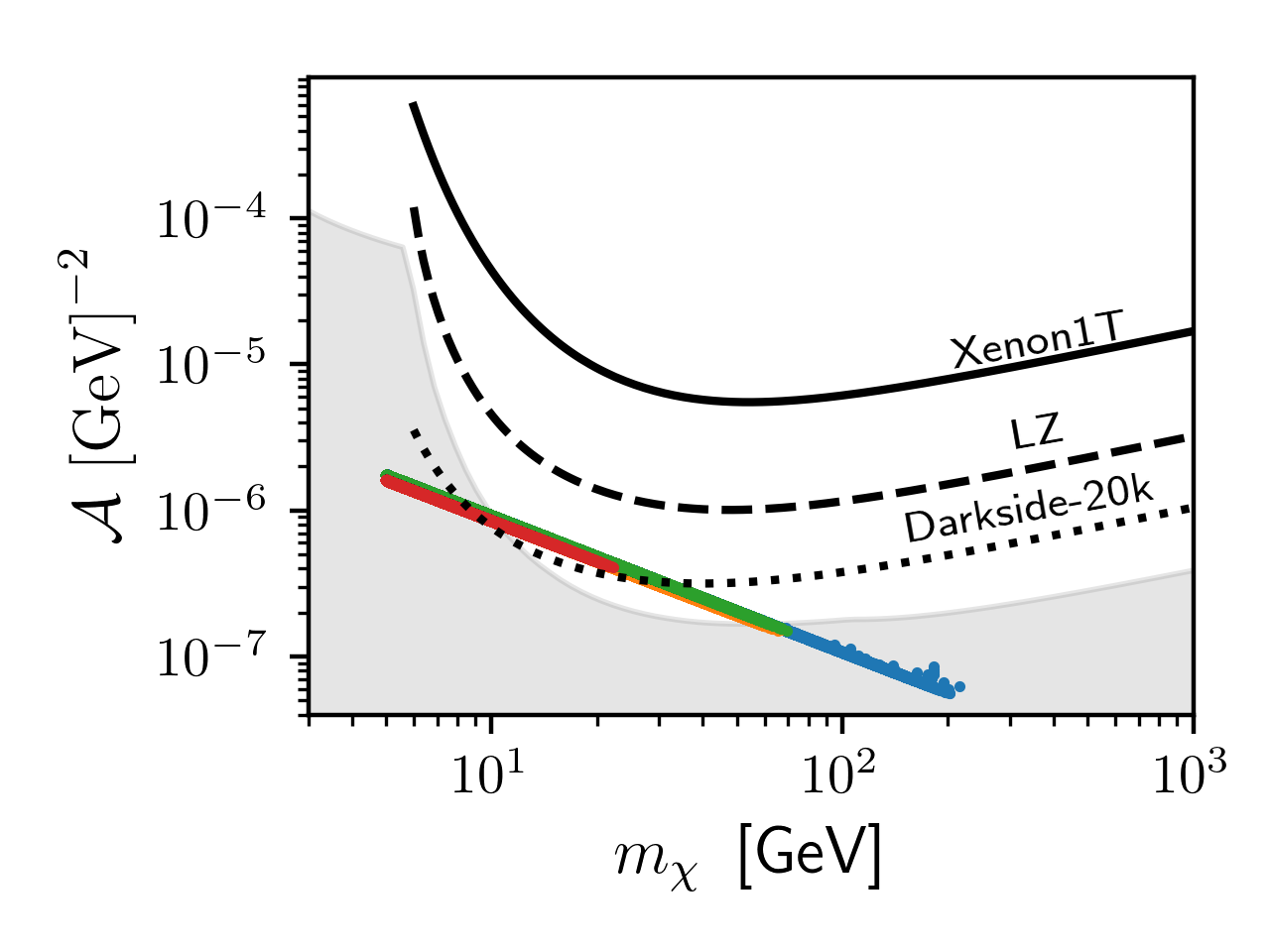}\hspace*{-0.3cm}
    \includegraphics[width=0.53\textwidth]{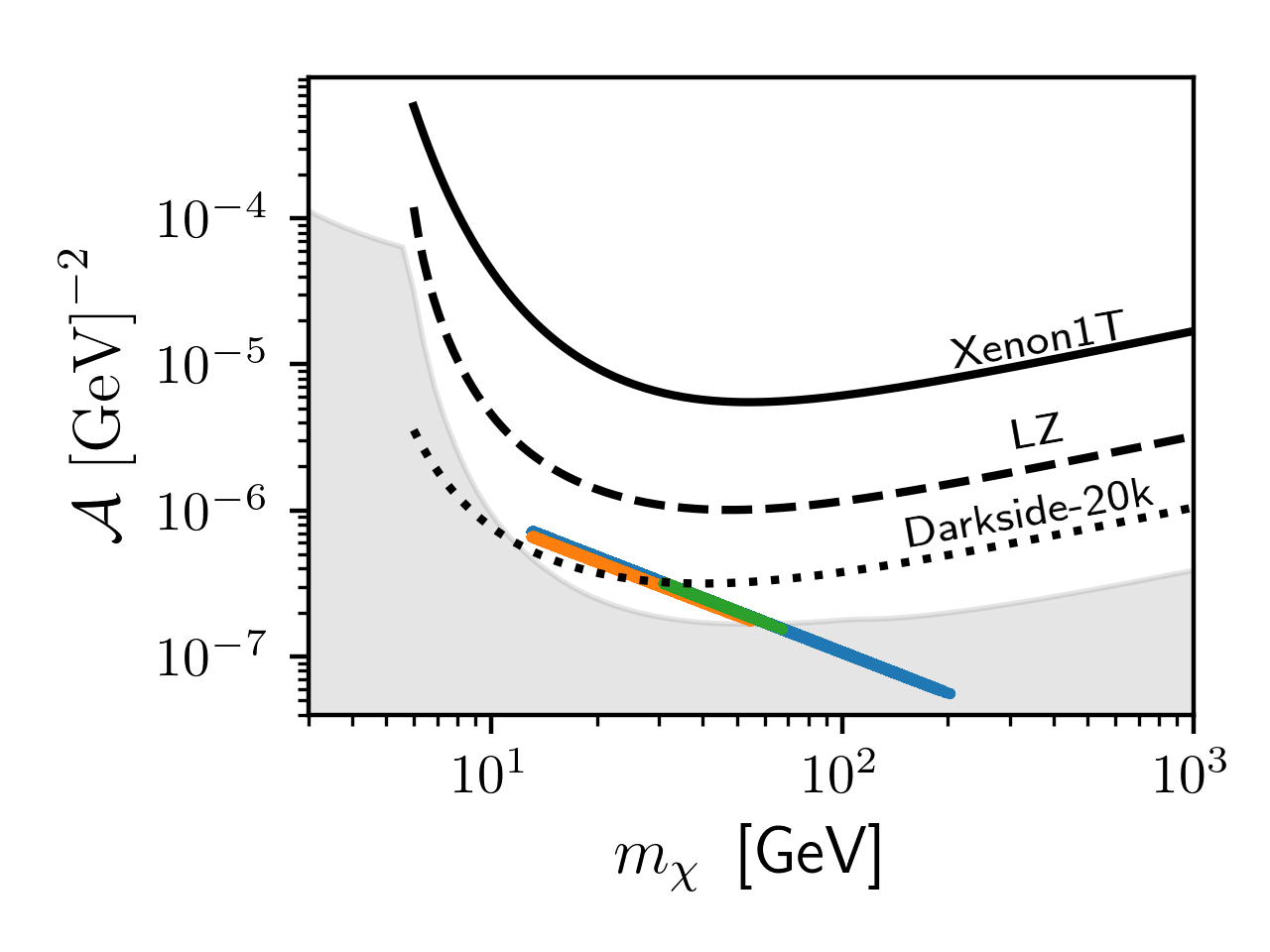}
    \caption{Theoretical predictions for the anapole coupling, $\mathcal{A}$, as a function of the DM mass, $\mchi$ for the four benchmark points: A1 (red points), A2 (green), B1 (orange), and B2 (blue). For comparison, we show the current exclusion line by Xenon1T \cite{Aprile:2018dbl} and the predicted reach of LZ \cite{Akerib:2015cja,Mount:2017qzi} and DarkSide-20k \cite{Aalseth:2017fik}. The shaded area represents the neutrino floor. The plot on the right-hand side incorporates LHC constraints, explained in more detail in Section\,\ref{sec:lhc}.}
    \label{fig:anapole}
\end{figure}

Given the range of DM masses that we consider in this study, the main constraint is due to Xenon1T results \cite{Aprile:2018dbl}, which we simulate using the prescription outlined in appendix A of Ref.~\cite{Kavanagh:2018xeh}, achieving good agreement. As we can see in Figure~\ref{fig:anapole}, the theoretical predictions for this model are beyond the reach of current experimental searches. We also show the reach of future direct detection experiments. The LZ detector, will employ 5.6 tons of liquid xenon with 1000 days exposure as outlined in \cite{Akerib:2015cja,Mount:2017qzi}. The DarkSide-20k experiment\cite{Aalseth:2017fik}, is an argon detector which will employ $20$ tons of fiducial mass for a duration of 10 years. We have assumed that the DarkSide collaboration will be able to achieve a threshold energy of $5$ keV, a reasonable assumption considering the results from DarkSide-50 \cite{Agnes:2018ves}. For reference we have also calculated the neutrino floor for anapole interactions in the $(\mathcal{A},\, \mchi)$ plane. We have used the prescription described in Ref.\,\cite{Billard:2013qya} and the expected neutrino fluxes from Refs.~\cite{Gaisser:2002jj,Battistoni:2005pd,Horiuchi:2008jz,Serenelli:2011py,2013arXiv1301.0365H,Gelmini:2018ogy}. It is clear that our model favourably lays in a region of parameter space that would be probed by a generation of experiments with multi-ton targets, that can probe near or even slightly beyond the neutrino floor. Spectral analysis with the neutrino background compounded with annual modulation data, could provide complete discrimination between model and the anapole moment which is both velocity and momentum dependent.

\begin{figure}[t!]
    \centering
    \hspace*{-0.3cm}\includegraphics[width=0.53\textwidth]{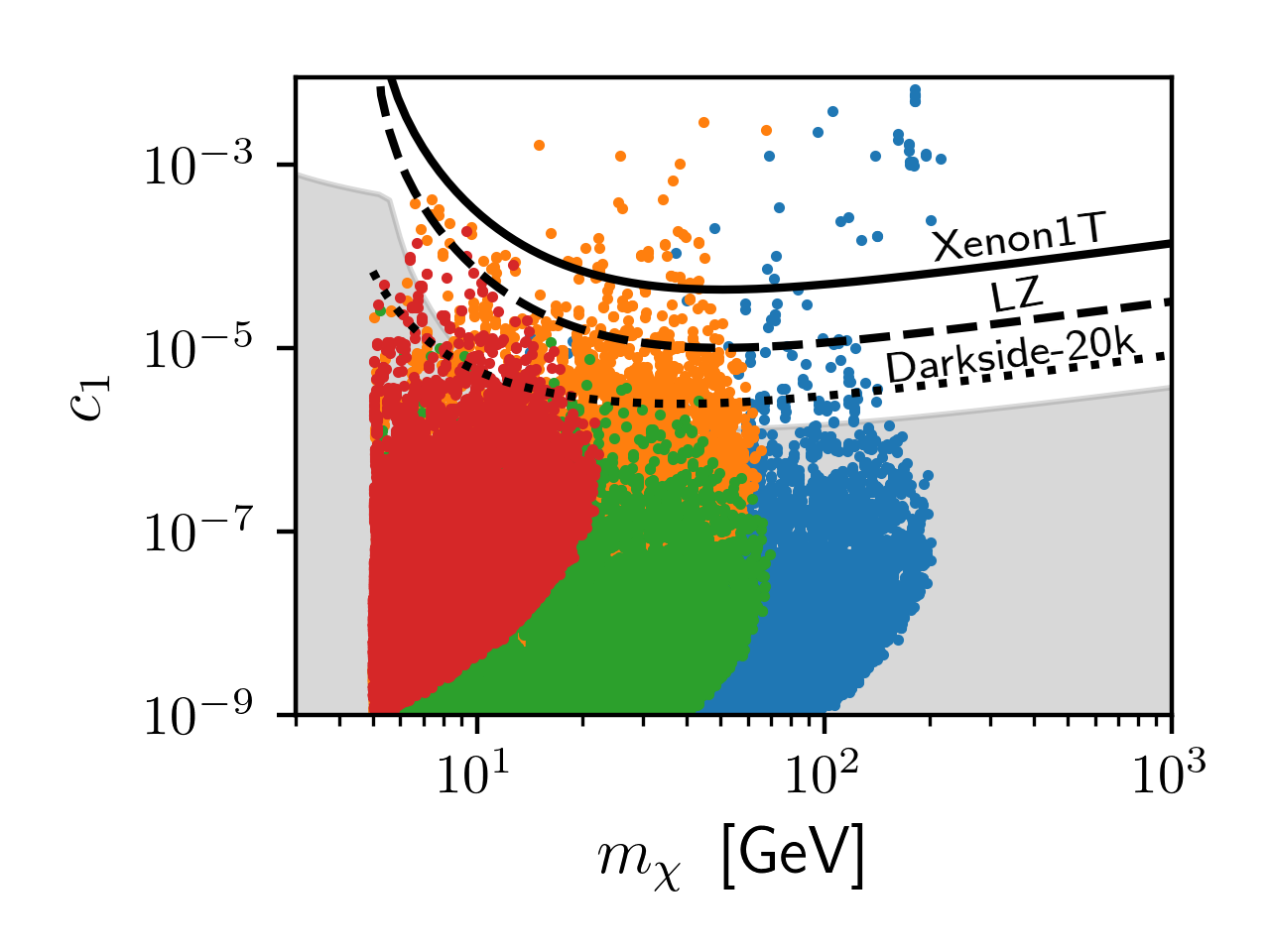}\hspace*{-0.5cm}
    \includegraphics[width=0.53\textwidth]{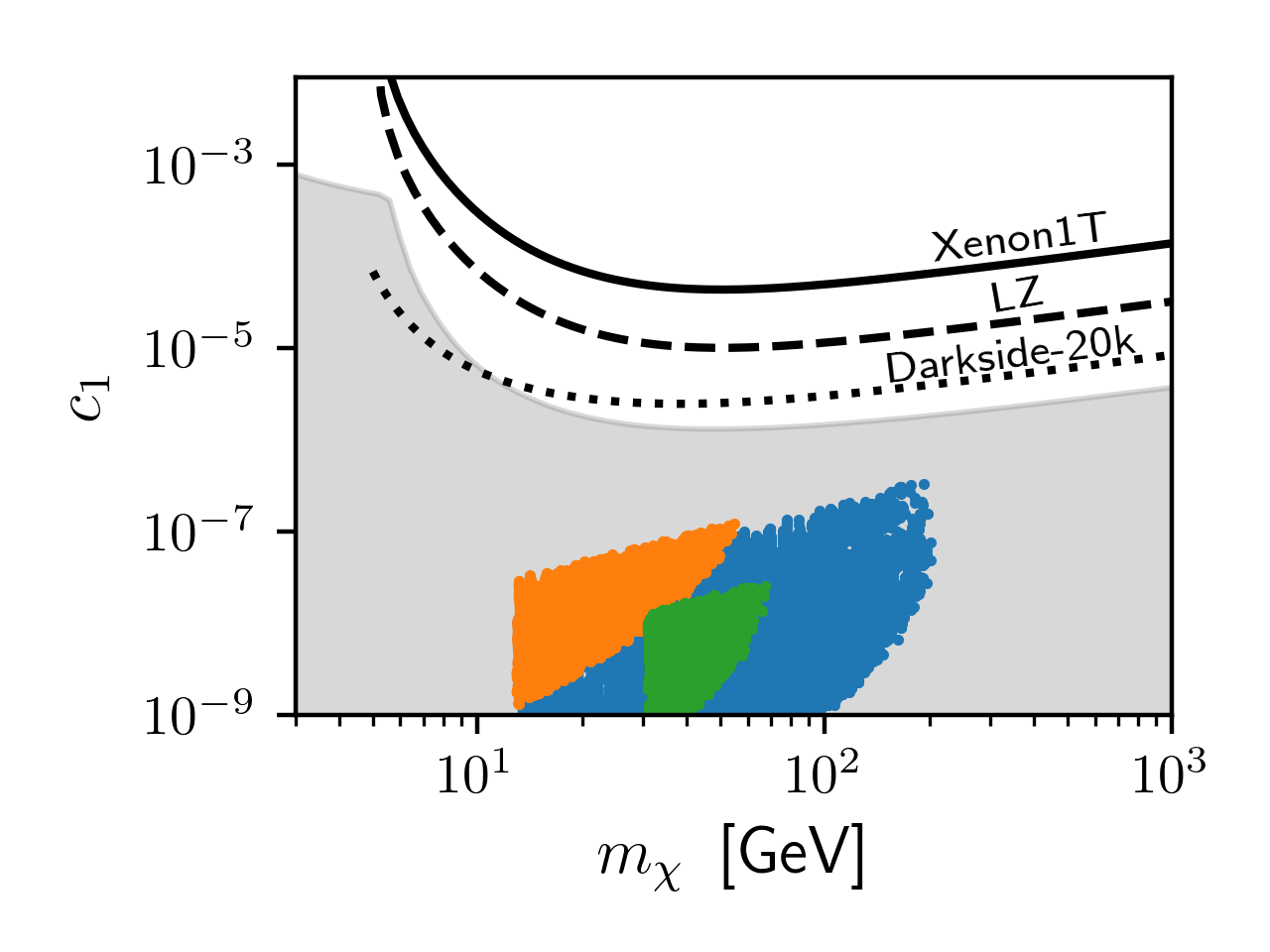}
    \caption{The same as in Figure~\ref{fig:anapole}, but for the spin-independent coupling, $c_1$, that originates from the twist-2 coupling.
    }
    \label{fig:twist2}
\end{figure}

For completeness, we have also calculated the effect on the total scattering cross section from aforementioned twist-2 operator and spin-dependent interaction. The former contribute to the spin-independent scattering cross section (operator ${\cal{O}}_1$) and can be sizeable if the new coupling to quarks is large or the colour mediator is very light. We have explicitly checked that once LHC constraints are included in the parameter space of the model, these terms are always subdominant to the anapole term discussed above. We represent in Figure~\ref{fig:twist2} the theoretical predictions for $c_1$ as a function of the DM mass from this contribution. For the spin-dependent interaction, we found that the predicted rate for our sampled parameter space is always sub-dominant.

Had we chosen to work with a Dirac fermion, the dipole and charge radius contributions should have been added. As it has been pointed out in Ref.~\cite{Ibarra:2015fqa}, the fairly large coupling to muons that is required to explain the flavour anomalies leads to effective DM couplings that are orders of magnitude higher than those coming from the tree level contribution, the most important being the charge-radius interaction. This we have checked, and in fact above $m_{\chi}\sim 10$ GeV, all our parameter points are excluded by Xenon1T. Below $m_{\chi}\sim 10$, the model is excluded by both LHC constraints and indirect detection bounds. Unlike in the Majorana case, the s-wave contribution to the thermal cross section $\sigmav$ is no longer helicity suppressed and hence excluded \cite{Leane:2018kjk}.

Our results suggest that future multi-ton direct detection experiments, such as DarkSide \cite{Aalseth:2017fik}, would be able to probe this model in the mass range $\mchi\sim 10-60$~GeV. It is very interesting to point out that many of the points in this DM mass range feature very heavy $\phi_q$ and therefore would be beyond the reach of collider searches. In a sense, future direct DM detection and the LHC complement each other to probe a large part of the model's parameter space.

\section{Conclusions}
\label{sec:conclusions}

In this article, we have studied a particle physics model that addresses the hints of lepton flavour universality violation observed by LHCb in $\bsmumu$ transitions, and that provides a solution to the dark matter problem. The scenario that we have analysed incorporates two new scalar fields and a Majorana fermion that provide one-loop contributions to $B$ meson decays.

The Majorana fermion is stable and can reproduce the observed DM relic abundance. We have studied the effect of new physics in flavour observables, for which $B_{s}-$mixing and $\bsmumu$ processes provide the most important constraints. In order to find an explanation for the $B$ anomalies and to reduce the $1.8\,\sigma$ tension between the predicted and measured mass difference in $B_{s}-$mixing, complex couplings are needed.  We have used results from the first global fit that takes into account this possibility. The combination of flavour bounds and constraints on the DM relic abundance leads to upper limits on the masses of the exotic states, and in general points towards a rather light DM candidate (with a mass $\mchi\lesssim 200$~GeV).

We have studied the signatures that this model would produce at the LHC. The dominant processes are the pair production of the coloured and leptonic scalars. For the former, the strongest exclusion limits are given by $\text{dijet} + \MET$ searches. For the latter, the final states are very clean, containing $1$ or $2$ leptons and missing energy. Both searches are complementary and exclude different regions of the parameter space, setting lower bounds on DM and mediator masses. The high-luminosity phase improves bounds coming from both searches, with dilepton being the most pronounced. The collider constraints are weakened when the $\lambda_{\mu}$ parameter is pushed towards the perturbative limit.

Finally, we have investigated how DM direct detection experiments constrain this model. Given the range of DM masses that we consider in this study, the main constraint is due to Xenon1T results. The small new couplings required by flavour constraints means that one-loop contributions to the DM-nucleon scattering cross section are generally larger than the tree level process. In particular, the dominant loop induced interaction is the anapole moment. We have shown that this model is not excluded by current data and could be probed by the next generation of experiments with multi-ton targets in the mass range $\mchi\sim 10-60$~GeV.

\vspace*{2ex}

\noindent
{\Large{\bf Acknowledgements}}

We would like to thank Alejandro Ibarra for useful discussions. DC and AC are grateful for the support from the Science and Technology Facilities Council (STFC). DC acknowledges the partial support of the Centro de Excelencia Severo Ochoa Program through the IFT-UAM/CSIC Associate programme.
PMR and JMM acknowledge support from the Spanish Research Agency (Agencia Estatal de Investigaci\'on) through the contract FPA2016-78022-P and IFT Centro de  Excelencia Severo Ochoa under grant SEV-2016-0597.

\bibliographystyle{JHEP-cerdeno}
\bibliography{references}

\end{document}